\documentclass[preprint,prd,aps,showkeys,nofootinbib]{revtex4}
\usepackage{graphicx}

\UseRawInputEncoding

\usepackage{bm}
\begin{document}


\title{Nuclear $0\nu2\beta$ decays in $B-L$ symmetric SUSY model and in TeV scale left-right symmetric model}

\author{Jin-Lei Yang$^{1,2}$}
\email{yangjinlei@itp.ac.cn}
\author{Chao-Hsi Chang$^{1,2}$}
\email{zhangzx@itp.ac.cn}
\author{Tai-Fu Feng$^{1,3}$}
\email{fengtf@hbu.edu.cn}

\affiliation{$^1$CAS Key Laboratory of Theoretical Physics, Institute of Theoretical Physics,
Chinese Academy of Sciences, Beijing 100190, China\\
$^2$School of Physical Sciences, University of Chinese Academy of
Sciences, Beijing 100049, China\\
$^3$Department of Physics, Hebei University, Key Laboratory of High-precision
Computation and Application of Quantum Field Theory of Hebei Province, Baoding, 071002, China}

\begin{abstract}

In this paper we take B-L supersymmetric standard model (B-LSSM) and TeV scale left-right symmetric model (LRSM) as two representations of the two kinds of new physics models to study the nuclear neutrinoless double beta decays ($0\nu2\beta$) so as to see the senses onto these two kinds of models when the decays are taken into account additionally. Within the parameter spaces allowed by all the exist experimental data, the decay half-life of the nucleus $^{76}$Ge and $^{136}$Xe, $T^{0\nu}_{1/2}$($^{76}$Ge, $^{136}$Xe), is precisely calculated and the results are presented properly. Based on the numerical results, we conclude that the room of LRSM type models for the foreseeable future experimental observations on the decays is greater than that of B-LSSM type models.

\end{abstract}


\maketitle

\section{Introduction\label{sec1}}

Tiny but nonzero neutrino masses explaining neutrino oscillation experiments~\cite{Tanabashi:2018oca}
are unambiguous evidences of new physics (NP) beyond the standard model (SM). It is because that in
SM there are only left-handed neutrinos, so that the neutrinos can acquire neither Dirac masses
nor Majorana masses. Hence to explore any mechanism inducing the tiny neutrino masses,
as well as relevant phenomenology, is an important direction to search for NP. The simple extension
of SM is to introduce three right-handed neutrinos in singlet of the gauge group SU(2) additionally, where
the neutrinos acquire Dirac masses, and to fit the neutrino oscillation and nuclear decay experiments
as well as astronomy observations, the corresponding Yukawa couplings of Higgs to the neutrinos are requested
so tiny as $\lesssim 10^{-12}$, that is quite unnatural.

However neutrino(s) may acquire masses naturally by introducing Majorana mass terms in extended SMs.
Once a Majorana mass term is introduced, certain interesting physics arise. One of the consequences
is that the lepton-number violation (LNV) processes, e.g. the nuclear neutrinoless double beta
decays ($0\nu2\beta$) etc may occur. Of them, the $0\nu2\beta$ decays are specially interesting, because
they may tell us sensitively some on the nature, whether Dirac~\cite{Dirac:1925jy} or
Majorana~\cite{Majorana:1937vz}, of the neutrinos. When the decays $0\nu2\beta$ are observed
in experiments, most likely, the neutrinos contain Majorana components.
Thus studying $0\nu2\beta$ decays is attracting special attentions.

Nowadays there are several experiments running to observe the $0\nu2\beta$ decays,
and the most stringent experimental bounds on the processes are obtained by
GERDA~\cite{Agostini:2017iyd,Guiseppe:2017rcr} and KamLAND-Zen~\cite{KamLAND-Zen:2016pfg,Albert:2017owj}.
They adopt suitable approaches and nucleus such as $^{76}$Ge and $^{136}$Xe respectively.
Now the latest experimental lower bound on the decay half-life given by GERDA experiments
is $T_{1/2}^{0\nu}>1.8\times10^{26}$ years ($90\%$ C.L.) for nucleus $^{76}$Ge~\cite{Agostini:2020xta}, and
in the near future the sensitivity can reach to $10^{28}$ years~\cite{talk}.
For the nucleus $^{136}$Xe, the most stringent lower bound on the decay half-life is $T_{1/2}^{0\nu}>1.07\times10^{26}$ years ($90\%$ C.L.)
given by KamLAND-Zen~\cite{KamLAND-Zen:2016pfg}, and the corresponding future sensitivity can reach to $2.4\times10^{27}$ years~\cite{Agostini:2020adk}. Moreover, underground experiments PANDAX, CDEX etc, which are originally designed for
searching for WIMP dark matter, are also planning to seek the $0\nu2\beta$ decays i.e. they may observe the $0\nu2\beta$ decays
with the sensitivity which at least will set a fresh lower bound.

In literature, there are a lot of theoretical analyses on the $0\nu2\beta$ decays.
The analyses are carried out generally by dividing the estimation of the $0\nu2\beta$ decays
into three `factors': one is at quark level to evaluate the amplitude for the `core' process $d+d\to u+u+e+e$ of
the decays; the second one is, from quark level to nucleon level, to involve the quark process into the relevant
nucleon one i.e. the `initial' quarks $d$ and $d$ involve into the two neutrons in the initial nucleus and
the `final' quarks $u$ and $u$ involve into the two protons in the final nucleus; the third one is, from nucleon level to nucleus
level, the relevant nucleons involve in the initial nucleus and the final nucleus properly. For the `core' process, in Ref.~\cite{Pas:1999fc} a general Lorentz-invariant effective Lagrangian is constructed by dimension-9 operators, and in Ref.~\cite{Gonzalez:2015ady,Liao:2020roy} the QCD corrections to all of these dimension-9 operators are calculated. In Ref.~\cite{Pas:2000vn} the short-range effects at quark level are considered, the analyses of the decay rates in the SM effective field theory are presented in Refs.~\cite{Cirigliano:2017djv,Cirigliano:2018hja,
Cirigliano:2018yza}, the $0\nu2\beta$ decay rates are derived in Refs.~\cite{Dekens:2020ttz}, the corresponding nuclear matrix elements (NME) and phase-space factors (PSF) for the second and the third factors of the decays i.e. from quark level to nucleon and nucleus levels, are considered in Refs.~\cite{Barea:2015kwa,Suhonen:1991sk,Caurier:2007xz,Simkovic:2007vu,Menendez:2008jp,Barea:2009zza,Rodriguez:2010mn,Suhonen:2012wd,
Barea:2013bz,Graf:2018ozy,Deppisch:2020ztt}, some theoretical predictions on the $0\nu2\beta$ for certain models are presented in Refs.~\cite{Deppisch:2012nb,LopezPavon:2012zg,Awasthi:2013ff,Asaka:2016zib,Pascoli:2013fiz}, and the theoretical analyses on the decays are reviewed in Refs.~\cite{DellOro:2016tmg,Engel:2016xgb,Dolinski:2019nrj}.

In this work, we are investigating the constrains from the $0\nu2\beta$ decays for the $B-L$ supersymmetric model (B-LSSM) and for the TeV scale left-right symmetric model (LRSM) comparatively. It is because that the two models are typical: both have LNV source but the mechanisms which give rise to the Majorana mass terms are different~\cite{Dev:2016dja,Patra:2015bga,Mitra:2016kov,Maiezza:2016ybz,Khalil:2008ps,
Elsayed:2011de,Khalil:2015naa,DelleRose:2017ukx,Yang:2018fvw,Yang:2018utw,Yang:2018guw,Yang:2019aao,Yang:2020ebs,
Yang:2020bmh}. In the B-LSSM, the tiny neutrino masses are acquired naturally through the so-called type-I seesaw mechanism which is proposed firstly by Weinberg~\cite{Weinberg:1979sa}. In the LRSM~\cite{Mohapatra:2006gs,Mohapatra:2005wg}, the tiny neutrino masses are acquired by both of type-I and type-II seesaw mechanisms, in addition, the new right-handed gauge bosons $W^\pm_R$ is introduced in this model, then both of left-handed and right-handed currents cause the $0\nu2\beta$
decays~\cite{Mohapatra:1979ia,Mohapatra:1981pm,Picciotto:1982qe,Hirsch:1996qw,Tello:2010am,Chakrabortty:2012mh,Barry:2013xxa,Dev:2013vxa,
Dev:2013oxa,Huang:2013kma,Dev:2014xea,Ahmed:2019vum,Borah:2015ufa,Ge:2015yqa,Awasthi:2015ota,Bambhaniya:2015ipg,Pritimita:2016fgr,Deppisch:2017vne}. As results, the computations of the decays are much more complicated in the LRSM than those in the B-LSSM. Hence these two models, being representatives of NP models, are typical for the $0\nu2\beta$ decays, and one may learn the mechanisms in the models well via analyzing the $0\nu2\beta$ decays comparatively.

In the study here, we will mainly focus on the first `factor' about the quark level i.e. the core process, which relates to the applied specific model closely. We will evaluate the Wilson coefficients of the operators relevant to the core process $d+d\to u+u+e+e$ etc on the models, whereas the estimation of the other two `factors', i.e. to evaluate `NME' and `PSF' etc, we will follow the literatures~\cite{Graf:2018ozy,Deppisch:2012nb,LopezPavon:2012zg,Awasthi:2013ff,Asaka:2016zib}. Respect to the `core' process $d+d\to u+u+e+e$, all of the contributions in the B-LSSM can be deduced directly quite well, while the contributions in LRSM cannot be so. As shown in Ref.~\cite{Kotila:2021xgw}, the calculations in the LRSM are much complicated and the interference effects are quite hard to be considered well. In this work, a new approximation, i.e. the momenta of the two involved quarks inside the initial or final nuclei is tried to be set equal, is made so as may reduce all contributions in the LRSM quite similar to the case of B-LSSM. Then the calculations in LRSM are simplified quite a lot and under the approximation the interference effects can be treated comparatively well. Finally, for comparison, we also present the results obtained by the traditional method~\cite{Dev:2014xea}.

The paper is organized as follows: In Sec.~\ref{sec2}, for B-LSSM, the seesaw mechanisms which give rise to the tiny neutrino masses, the heavy neutral leptons as well, the relevant interactions etc, the calculations of the $0\nu2\beta$ decay half-lifes of the nuclei are given. Similarly, in Sec.~\ref{sec3}, for LRSM, the seesaw mechanisms which give rise to the tiny neutrino masses, as well the heavy neutral leptons, the relevant interactions and the calculations of the $0\nu2\beta$ decay half-lifes of the nuclei are given. In Sec.~\ref{sec4A} and~\ref{sec4B} the numerical results for B-LSSM and LRSM are presented respectively. Finally, in Sec.~\ref{sec5} brief discussions and conclusions are given. In the appendix, the needed QCD corrections to the effective Lagrangian which contains the dimension-9 operators in the region from the energy scale $\mu\simeq M_W$ to the energy scale $\mu\simeq 1.0\;$GeV are collected.

\section{The B-LSSM for $0\nu2\beta$ decays \label{sec2}}

In the B-LSSM, the local gauge group is $SU(3)_C\bigotimes SU(2)_L\bigotimes U(1)_Y\bigotimes U(1)_{B-L}$,
where $B$, $L$ denote the baryon number and lepton number respectively, and the details about
the gauge fields, their breaking, extra lepton and Higgs fields etc can be found in
Refs.~\cite{Khalil:2008ps,Elsayed:2011de,Khalil:2015naa,DelleRose:2017ukx,Yang:2018fvw,Yang:2018utw,
Yang:2018guw,Yang:2019aao,Yang:2020ebs,Yang:2020bmh}. In this model, tiny neutrino masses are acquired by the so-called type-I seesaw mechanism, when the $U(1)_{B-L}$ symmetry is broken spontaneously by the two $U(1)_{B-L}$ singlet scalars (Higgs). The mass matrix for neutrinos and neutral heavy leptons in the model can be expressed as
\begin{eqnarray}
&&\left(\begin{array}{c}0,\;\;\;\;\;\;\;M_D^{\rm T}\\M_D,\;\;\;\;M_R\end{array}\right),\label{eq2}
\end{eqnarray}
and the mass matrix can be diagonalized in terms of a unitary matrix $U_\nu$ as follows:
\begin{eqnarray}
&&U_\nu^{\rm T}\left(\begin{array}{c}0,\;\;\;\;\;\;\;M_D^{\rm T}\\M_D,\;\;\;\;M_R\end{array}\right)U_\nu=
\left(\begin{array}{c}\hat m_\nu,\;\;\;\;0\\0,\;\;\;\;\hat M_N\end{array}\right),
\end{eqnarray}
where the neutrino masses $\hat m_\nu={\rm diag}(m_{\nu_1},m_{\nu_2},m_{\nu_3})$, the masses of the heavy neutral leptons $\hat M_N={\rm diag}(M_{N_1},M_{N_2},M_{N_3})$ and $U_\nu$ is a matrix of $6\times6$ which can be rewritten as
\begin{eqnarray}
&&U_\nu=\left(\begin{array}{c}U\quad S\\T\quad V\end{array}\right),\label{eq6}
\end{eqnarray}
where $U,S,T,V$ are matrices of $3\times3$.

The interactions, being applied later on, in the model are
\begin{eqnarray}
&&\mathcal{L}_I=\frac{{\rm i}g_2}{\sqrt2}\sum_{j=1}^3\Big[U_{ij}\bar e_i\gamma^\mu P_L\nu_j W_{L,\mu}^-+S_{ij}\bar e_i\gamma^\mu P_L N_j W_{L,\mu}^-+h.c\Big],
\end{eqnarray}
where $\nu,\;N$ are the four-component fermion fields of the light and heavy neutral leptons respectively.

\begin{figure}
\setlength{\unitlength}{1mm}
\centering
\includegraphics[width=5in]{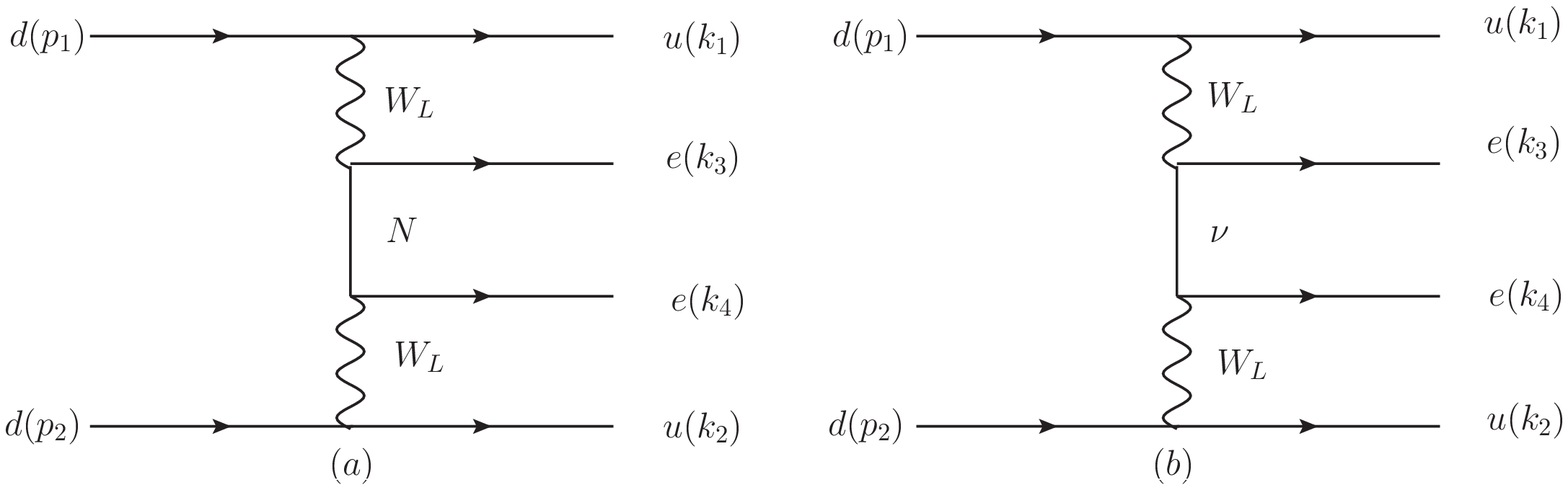}
\vspace{0cm}
\caption[]{The dominant Feynman diagrams for the $0\nu2\beta$ decays in the B-LSSM Model. (a) The contributions from the heavy neutral lepton exchanges, (b) The contributions from the light neutrino exchangs.}
\label{FeynmanBLSSM}
\end{figure}
The Feynman diagrams which response for the dominated contributions to the $0\nu2\beta$
decays in the B-LSSM are plotted in Fig.~\ref{FeynmanBLSSM}. Note here that the contributions from charged Higgs exchange(s) are ignored safely as they are highly suppressed by the charged Higgs masses and Yukawa couplings, thus in Fig.~\ref{FeynmanBLSSM} the charged Higgs exchange do not appear at all. To evaluate the contributions corresponding to the Feynman diagrams for the $0\nu2\beta$ decays, and to consider the roles of the neutral leptons (light neutrinos and heavy neutral leptons) in the Feynman diagrams, the useful formulae are collected below so as to deal with the neutrino propagator sandwiched by various chiral project operators $P_{L,R}=\frac{1}{2}(1\mp\gamma_5)$:
\begin{eqnarray}
&&P_L\frac{k\!\!\!/+m}{k^2-m^2}P_L=\frac{m}{k^2-m^2}P_L,\;\;P_R\frac{k\!\!\!/+m}{k^2-m^2}P_R=\frac{m}{k^2-m^2}P_R,
\nonumber\\
&&P_L\frac{k\!\!\!/+m}{k^2-m^2}P_R=\frac{k\!\!\!/}{k^2-m^2}P_R,\;\;P_R\frac{k\!\!\!/+m}{k^2-m^2}P_L=
\frac{k\!\!\!/}{k^2-m^2}P_L,\label{a1}
\end{eqnarray}
and the propagator
\begin{eqnarray}
&&\frac{k\!\!\!/+m}{k^2-m^2}\simeq\left\{\begin{array}{l}
-\frac{1}{m}\;,\;\;\; m^2\gg k^2\\
\frac{k\!\!\!/+m}{k^2}\;,\;\;\; m^2\ll k^2\\
\end{array}\right.,\label{a2}
\end{eqnarray}

Relating to the exchanges of the heavy neutral leptons (the virtual neutral lepton momentum $k$ has $|k|\simeq 0.10\;{\rm GeV}\ll M_{N_i}$) for the decays, from Fig.~\ref{FeynmanBLSSM}~(a) the effective Lagrangian at the energy scale $\mu\simeq M_{W_L}$ can be read out as
\begin{eqnarray}
&&\frac{2m_p}{G_F^2\cos\theta_C^2}\mathcal{L}_{\rm eff}^{\rm BL}(N)=\sum_i\frac{2m_p}{M_{N_i}}(S_{1i})^2[4(\bar u \gamma_{\mu} P_L d)
(\bar u \gamma^{\mu} P_L d)\bar e P_R e^c]\equiv C_{3R}^{LL}(N)\mathcal{O}_{3R}^{LL},\nonumber\\
&&C_{3R}^{LL}(N)=\sum_i\frac{2m_p}{M_{N_i}}(S_{1i})^2,\;\mathcal{O}_{3Z}^{XY}=8(\bar u \gamma_{\mu} P_X d)
(\bar u \gamma^{\mu} P_Y d)\bar e P_Z e^c,\label{eq45}
\end{eqnarray}
where $X,Y,Z=L,R$, $\theta_C$ is the Cabibbo angle, $m_p$ is proton mass introduced for normalization of the effective Lagrangian, and $S_{1i}$ is the matrix elements in Eq.~(\ref{eq6}).

Since the nuclear $0\nu2\beta$ decays take place at the energy scale about $\mu\approx 0.10\;$GeV, obviously we need to consider the QCD corrections for the effective Lagrangian obtained at the energy scale $\mu\simeq M_{W_L}$ in Eq.~(\ref{eq45}) i.e. to evolute the effective Lagrangian in terms of renormalization group equation (RGE) method from the energy scale $\mu\simeq M_{W_L}$ to that $\mu\simeq 1.0\;$GeV first, where the corrections are in perturbative QCD (pQCD) region. Thus, for completeness, the QCD corrections to all of the possible dimension-9 operators which may contribute to the nuclear $0\nu2\beta$ decays, are calculated by the RGE method, and the details of computations are collected in the appendix. Whereas the QCD corrections in the energy scale region $\mu \simeq 1.0\;$GeV$\sim \mu \simeq 0.10\;$GeV, being in non-perturbative QCD region, we take into account them by inputting the experimental measurements for the relevant current matrix elements of nucleons, which emerge when calculating the amplitude based on the effective Lagrangian at $\mu\simeq 0.10\;$GeV.

For the decays when considering contributions from the neutrino ($m_{\nu_i}\ll |k|$) exchanges as described by the Feynman diagram Fig.~\ref{FeynmanBLSSM}~(b), and the interferences between the light neutrinos' and heavy neutral leptons' contributions, to derive the effective Lagrangian for the neutrino contributions is better at the energy scale $\mu\simeq 1.0\;{\rm GeV}$ as heavy neutral leptons too. At this energy scale the effective Lagrangian can be written down according to the Feynman diagram Fig.~\ref{FeynmanBLSSM}~(b) as below:
\begin{eqnarray}
&&\frac{2m_p}{G_F^2\cos\theta_C^2}\mathcal{L}_{\rm eff}^{\rm BL}(\nu)=\frac{m_{\nu_i}}{m_e}(U_{1i})^2\cdot\frac{2m_p m_e}{-k^2}\mathcal{O}_{3R}^{LL}\equiv C_{3R}^{LL}(\nu)\frac{2m_p m_e}{-k^2}\mathcal{O}_{3R}^{LL},\nonumber\\
&&C_{3R}^{LL}(\nu)=\frac{m_{\nu_i}}{m_e}(U_{1i})^2.\label{eq42}
\end{eqnarray}
Since the light neutrino exchange is of long range, the QCD corrections in the region $\mu\simeq 1.0\;$GeV$\sim \mu\simeq 0.10\;$GeV to the coefficients in Eq.~(\ref{eq42}) may be involved just as that in the case done as in the above for the heavy neutral lepton case, via inputting the experimental measurements for the relevant current matrix elements of nucleons, which emerge when calculating the matrix elements in the amplitudes at $\mu\approx 0.10\;$GeV.

To evaluate the half-life $T_{1/2}^{0\nu}\equiv\ln 2/\Gamma$ of the $0\nu2\beta$ decays, the contributions from the heavy neutral leptons (Fig~\ref{FeynmanBLSSM}~(a)) and those from the neutrinos (Fig~\ref{FeynmanBLSSM}~(b)) should be summed up for the amplitudes. The half-life $T_{1/2}^{0\nu}\equiv\ln 2/\Gamma$ of the $0\nu2\beta$ decays can be written as~\cite{Deppisch:2020ztt}
\begin{eqnarray}
&&\frac{1}{T_{1/2}^{0\nu}}=G^{0\nu}|M^{0\nu}|^2\Big|\frac{m_{ee}^{\rm BL}}{m_e}\Big|^2,
\end{eqnarray}
where $G^{0\nu}=2.36\times10^{-15}\;(14.56\times10^{-15})\;{\rm years}^{-1}$~\cite{Deppisch:2020ztt} for $^{76}{\rm Ge}\;(^{136}{\rm Xe})$ is the PSF, $M^{0\nu}=-6.64\pm1.06$ $(-3.60\pm0.58)$~\cite{Barea:2015kwa,Deppisch:2020ztt} for $^{76}{\rm Ge}\;(^{136}{\rm Xe})$\footnote{The NMEs adopted here are obtained within the framework of the microscopic interacting meson model for nuclei~\cite{Deppisch:2020ztt}, and the uncertainties for NMEs calculations from the various nuclear structure models are quite wild, here we remind only that the NMEs obtained by various approaches are varying by a factor of $(2\sim3)$ roughly~\cite{Kotila:2021gwk}.} is the NME corresponding to the long range contributions which is defined as
\begin{eqnarray}
&&M^{0\nu}\equiv\langle{\mathcal O}_F^+|\frac{2m_p m_e}{-k^2}[4(\bar u \gamma_{\mu} P_L d)
(\bar u \gamma^{\mu} P_L d)]|\mathcal O_I^+\rangle\label{eqNME}
\end{eqnarray}
with $|\mathcal O_I^+\rangle$, $\langle\mathcal O_F^+|$ denoting the initial and final nuclear states respectively. Note that in Eq.~(\ref{eqNME}) that the factor $\frac{2m_p m_e}{-k^2}$ in Eq.~(\ref{eq42}) is absorbed into the so-called `neutrino potential' which is used to compute the long range NME. And
\begin{eqnarray}
&&m_{ee}^{\rm BL}\equiv U_3^{XX}C_{3R}^{LL}(N)\frac{M_3^{XX}(N)}{M^{0\nu}}+C_{3R}^{LL}(\nu),\label{meeBL}
\end{eqnarray}
where $U_3^{XX}$ is the QCD running factor from $\mu\simeq M_{W_L}$ to $\mu\simeq 1.0\;$GeV (the numerical result of $U_3^{XX}$ can be found in Eq.~(\ref{48})), $M_3^{XX}(N)=-200\pm56\;(-111\pm31.08)$~\cite{Barea:2015kwa,Deppisch:2020ztt} for $^{76}{\rm Ge}\;(^{136}{\rm Xe})$ is the NME corresponding to short range contributions which is defined as
\begin{eqnarray}
&&M_3^{XX}(N)\equiv\langle{\mathcal O}_F^+|[4(\bar u \gamma_{\mu} P_X d)
(\bar u \gamma^{\mu} P_X d)]|\mathcal O_I^+\rangle.
\end{eqnarray}

\section{The LRSM\label{sec3}}

For the model LRSM, the gauge fields are $SU(3)_C\bigotimes SU(2)_L\bigotimes SU(2)_R\bigotimes U(1)_{B-L}$,
and the details about the gauge fields and their breaking can be found in Refs.~\cite{Dev:2016dja,Patra:2015bga,Mitra:2016kov,Maiezza:2016ybz}.
In this model, the tiny neutrino masses are obtained by both of type-I and type-II seesaw mechanisms
due to introducing the right-handed neutral leptons and two triplet Higgs (scalars) accordingly. In the model, the mass matrix for the neutral leptons generally is written as
\begin{eqnarray}
&&\left(\begin{array}{c}M_L,\;\;\;\;M_D^{\rm T}\\M_D,\;\;\;\;M_R\end{array}\right) \label{eq11}
\end{eqnarray}
and the mass matrix Eq.~(\ref{eq11})\footnote{The matrix Eq.~(\ref{eq11}) with $M_L=0$ indicates the masses are acquired by type-I seesaw mechanism; it with $M_D=0$ indicates the masses are acquired
by type-II seesaw mechanism; it in general feature indicates the masses are acquired
by type-I+II seesaw mechanism.} can be diagonalized in terms of a unitary matrix $U_\nu$, whereas
the matrix $U_\nu$ can be expressed similarly as that in the case of the B-LSSM Eq.~(\ref{eq6}).

For the model LRSM, if the left-right symmetry is not broken manifestly but
spontaneously, i.e. $g_L=g_R\equiv g_2$ and as one consequence, the mass terms of $W$ bosons can be written as
\begin{eqnarray}
&&\mathcal{L}_{M_W}=\frac{g_2^2}{4}\left(\begin{array}{c}W_{L}^+,\;\;W_{R}^+\end{array}\right)
\left(\begin{array}{c}v_1^2+v_2^2+2v_L^2,\quad 2v_1v_2\\ 2v_1v_2,\quad v_1^2+v_2^2+2v_R^2\end{array}\right)
\left(\begin{array}{c}W_{L}^-\\ W_{R}^-\end{array}\right),
\end{eqnarray}
where $v_1, v_2, v_L, v_R$ ($v_L\ll v_R$) are the VEVs of new scalars (Higgs) in the LRSM. Then the physical masses of the $W$ bosons can be obtained~\cite{Deppisch:2017vne}
\begin{eqnarray}
&&M_{W_1}\simeq\frac{g_2}{2}(v_1^2+v_2^2)^{1/2}, \;\;\;\;\; M_{W_2}\simeq\frac{g_2}{\sqrt2}v_R.
\end{eqnarray}
The mass eigenstates $W_{1,2}^\pm$ are related to the interaction eigenstates $W_{L,R}^\pm$ by $\zeta$
\begin{eqnarray}
&&\left(\begin{array}{c}W_1^\pm\\ W_2^\pm\end{array}\right)=\left(\begin{array}{c}\cos \zeta,\;\;\sin\zeta\\ -\sin\zeta,\;\;\cos\zeta\end{array}\right)\left(\begin{array}{c}W_{L}^\pm\\ W_{R}^\pm\end{array}\right),
\end{eqnarray}
where $\tan2\zeta=\frac{2v_1v_2}{v_R^2-v_L^2}$.

The interactions, being applied later on, in the model are
\begin{eqnarray}
&&\mathcal{L}_I^{LRSM}=\frac{{\rm i}g_2}{\sqrt2}\sum_{j=1}^3\Big[\bar e_i(\cos\zeta U_{ij}\gamma^\mu P_L+\sin\zeta T^*_{ij}\gamma^\mu P_R)\nu_{j}W_{1,\mu}^-\nonumber\\
&&\qquad\qquad\;\;\;+\bar e_i(\cos\zeta T^*_{ij}\gamma^\mu P_R-\sin\zeta U_{ij}\gamma^\mu P_L)\nu_{j}W_{2,\mu}^-\nonumber\\
&&\qquad\qquad\;\;\;+\bar e_i(\cos\zeta S_{ij}\gamma^\mu P_L+\sin\zeta V^*_{ij}\gamma^\mu P_R)N_{j}W_{1,\mu}^-\nonumber\\
&&\qquad\qquad\;\;\;+\bar e_i(\cos\zeta V^*_{ij}\gamma^\mu P_R-\sin\zeta S_{ij}\gamma^\mu P_L)N_{j}W_{2,\mu}^-\nonumber\\
&&\qquad\qquad\;\;\;+\bar u(\cos\zeta\gamma^\mu P_L+\sin\zeta\gamma^\mu P_R) dW_{1,\mu}^-\nonumber\\
&&\qquad\qquad\;\;\;+\bar u(\cos\zeta\gamma^\mu P_R-\sin\zeta\gamma^\mu P_L) dW_{2,\mu}^-+h.c\Big],
\end{eqnarray}
where the definitions for $U,\;S,\;T,\;V,\;\nu,\;N$ are same as the ones in the B-LSSM.

\begin{figure}
\setlength{\unitlength}{1mm}
\centering
\includegraphics[width=5in]{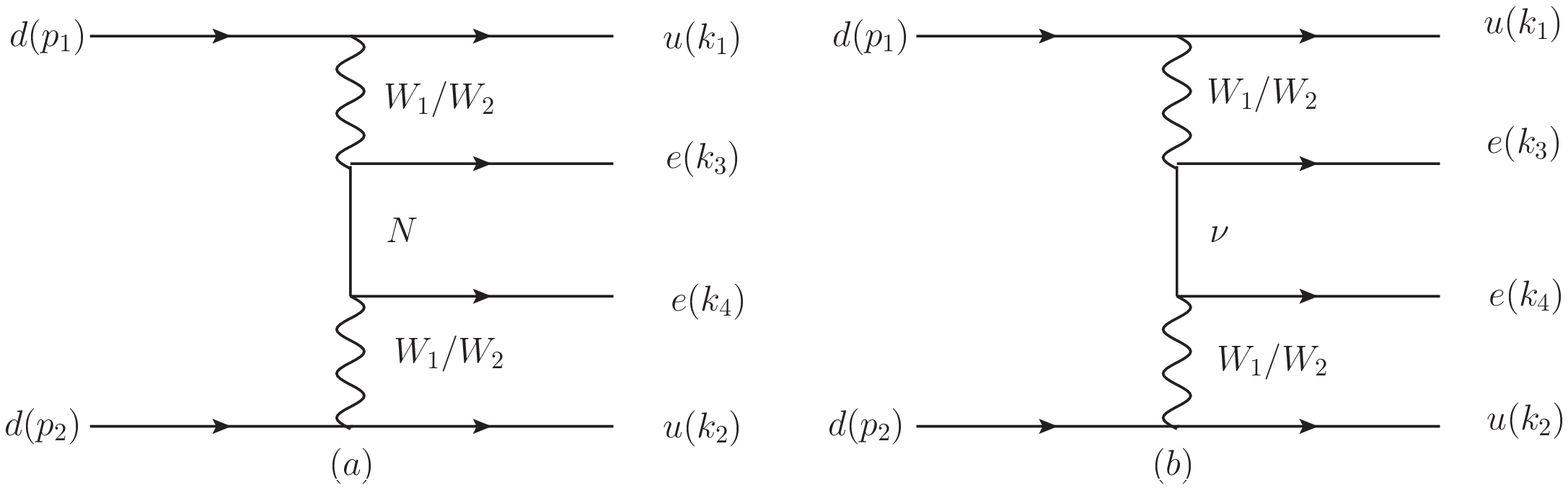}
\vspace{0cm}
\caption[]{The dominant Feynman diagrams for the $0\nu2\beta$ decays in the LRSM. (a) The contributions from the heavy neutral lepton exchanges, (b) The contributions from the light neutrino exchanges.}
\label{FeynmanLRSM}
\end{figure}

In the LRSM, the dominant contributions to the $0\nu2\beta$ decays are represented by Feynman diagrams Fig.~\ref{FeynmanLRSM}.
In comparison with those of the B-LSSM, the contributions from the Higgs exchanges can be ignored so in Fig.~\ref{FeynmanLRSM} there is no Higgs exchange at all, but besides the left-handed gauge boson components $W_L^\pm$, there are right-handed components $W_R^\pm$ in $W_1^\pm$ and $W_2^\pm$ gauge bosons. Therefore the situation in determining the effective Lagrangian for the $0\nu2\beta$ decays is different and comparatively complicated than that for the B-LSSM. In the case for the contributions from the heavy neutral lepton exchanges shown in Fig.~\ref{FeynmanLRSM} (a), considering the fact that the heavy neutral leptons propagator $\frac{M_{N_i}+ k\!\!\!/}{k^2-M_{N_i}^2}\approx \frac{-1}{M_{N_i}}$ ($M_{N_i}\geq M_{W_1}$) and $\sin\zeta, S_{1i}, T_{1i}\ll 1.0$ in the decays, by using Eqs.~(\ref{a1},~\ref{a2}) the effective Lagrangian at the energy scale $\mu\simeq M_{W_1}$ ($W_1$ is the lighter one boson between $W_{1,2}$) can be written down as follows
\begin{eqnarray}
&&\frac{2m_p}{G_F^2\cos\theta_C^2}\mathcal{L}_{\rm eff}^{\rm LR}(N)=C_{3R}^{LL}(N)\mathcal{O}_{3R}^{LL}+
C_{3L}^{LL}(N)\mathcal{O}_{3L}^{LL}+C_{3L}^{RL}(N)\mathcal{O}_{3L}^{RL}+C_{3L}^{RR}(N)\mathcal{O}_{3L}^{RR}\,;\nonumber\\
&&C_{3R}^{LL}(N)=\frac{2m_p}{M_{N_i}}\cos^4\zeta S^{2}_{1i},\;\;C_{3L}^{LL}(N)=\frac{2m_p}{M_{N_i}}\cos^2\zeta\sin^2\zeta V^{*2}_{1i}.\nonumber\\
&&C_{3L}^{RL}(N)=\frac{2m_p}{M_{N_i}}\cos^3\zeta\sin\zeta V^{*2}_{1i}(\frac{M_{W_L}}{M_{W_R}})^2,\;\;C_{3L}^{RR}(N)=\frac{2m_p}{M_{N_i}}\cos^4\zeta V^{*2}_{1i}(\frac{M_{W_L}}{M_{W_R}})^4.\label{eq47}
\end{eqnarray}
In literatures, due to small $\zeta$, the contributions corresponding to $C_{3L}^{RL}(N)$, $C_{3L}^{LL}(N)$ in Eq.~(\ref{eq47}) are neglected. However, since $\tan2\zeta=\frac{2v_1v_2}{v_R^2-v_L^2}$, i.e. $\zeta\approx xM_{W_1}^2/M_{W_2}^2$ and when $x\equiv v_2/v_1>0.02$~\cite{Bertolini:2014sua}, the terms with $C_{3L}^{RL}(N)$ and $C_{3L}^{LL}(N)$ can also make essential contributions compared with the terms with $C_{3L}^{RR}(N)$, $C_{3R}^{LL}(N)$. Thus in this work, when evaluating the $0\nu2\beta$ decays we would like to keep the contributions from the terms of $C_{3L}^{RL}(N)$, $C_{3L}^{LL}(N)$, and consider the QCD corrections to the effective Lagrangian Eq.~(\ref{eq47}) in the similar way as that in the B-LSSM.

As the next step, when considering the contributions from the light neutrino exchanges as Fig.~\ref{FeynmanLRSM} (b), owing to the fact that $W^\pm_{1,2}$ contain both components $W^\pm_{L,R}$ in LRSM, according to Eqs.(\ref{a1},\ref{a2}), the light neutrino propagators with chiral project operators $P_{L,R}\frac{m_{\nu_i}+ k\!\!\!/}{k^2-m_{\nu_i}^2}P_{L,R}$ or $P_{L,R}\frac{m_{\nu_i}+ k\!\!\!/}{k^2-m_{\nu_i}^2}P_{R,L}$, as central factors, finally contribute the factors as $\frac{m_{\nu_{i}}}{k^2}P_{L,R}$ or $\frac{k\!\!\!/}{k^2}P_{L,R}$ respectively to the results. Namely the factors $\frac{k\!\!\!/}{k^2}P_{L,R}$ are new and substantial, and in B-LSSM they do not appear at all. Moreover when the contributions from the `higher order' terms for $\sin\zeta,\;S_{1i},\;T_{1i}$, such as those small terms proportional to $\sin^2\zeta,\;S_{1i}^2,\;T_{1i}^2$ etc, are ignored, then the effective Lagrangian at the energy scale $\mu\simeq 1.0\;{\rm GeV}$ may be read out from Fig.~\ref{FeynmanLRSM}(b) as
\begin{eqnarray}
&&\frac{2m_p}{G_F^2\cos\theta_C^2}\mathcal{L}_{\rm eff}^{\rm LR}(\nu)=\cos^4\zeta \frac{U_{1i}^2m_{\nu i}}{m_e}\frac{2m_p m_e}{-k^2}\mathcal{O}_{3R}^{LL}+\nonumber\\
&&\qquad\qquad\qquad\qquad\quad \cos^3\zeta\sin\zeta U_{1i}T_{1i}^*[4(\bar u \gamma_{\mu} P_L d)
(\bar u \gamma_{\nu} P_L d)\bar e \gamma^{\mu}\frac{2m_pk\!\!\!/}{k^2}\gamma^{\nu} e^c]+\nonumber\\
&&\qquad\qquad\qquad\qquad\quad \cos^4\zeta U_{1i}T_{1i}^*\frac{M_{W_L}^2}{M_{W_R}^2}[4(\bar u \gamma_{\mu} P_R d)
(\bar u \gamma_{\nu} P_L d)\bar e \gamma^{\mu}\frac{2m_pk\!\!\!/}{k^2}\gamma^{\nu}P_R e^c].\label{eq15}
\end{eqnarray}
The QCD corrections in the energy scale region $\mu \simeq 1.0\;$GeV to $\mu \simeq 0.10\;$GeV, being of non-perturbative QCD, are taken into account by inputting in the experimental measurements for the relevant current matrix elements of nucleons, which emerge at the effective Lagrangian at $\mu\simeq 0.10\;$GeV.

In Refs.~\cite{Mohapatra:1979ia,Mohapatra:1981pm,Picciotto:1982qe,Hirsch:1996qw,Tello:2010am,Chakrabortty:2012mh,
Barry:2013xxa,Dev:2013vxa,Dev:2013oxa,Huang:2013kma,Dev:2014xea,Ahmed:2019vum}, the second term and the third term of Eq.~(\ref{eq15}) are defined as $\eta$, $\lambda$ respectively. Extracting the factors
\begin{eqnarray}
&&C_\eta=\cos^3\zeta\sin\zeta U_{1i}T_{1i}^*,\;\;C_\lambda=\cos^4\zeta U_{1i}T_{1i}^*\frac{M_{W_1}^2}{M_{W_2}^2},
\end{eqnarray}
then the operators $[4(\bar u \gamma_{\mu} P_L d) (\bar u \gamma_{\nu} P_L d)\bar e \gamma^{\mu}\frac{2m_pk\!\!\!/}{k^2}\gamma^{\nu} e^c]$, $[4(\bar u \gamma_{\mu} P_R d) (\bar u \gamma_{\nu} P_L d)\bar e \gamma^{\mu}\frac{2m_pk\!\!\!/}{k^2}\gamma^{\nu}P_R e^c]$ are attributed to the calculations of NME and PSF.

Whereas when calculating the NMEs and PSF, the interference effects among the contributions, especially to consider the contributions from the factors $\frac{k\!\!\!/}{k^2}P_{L,R}$ for the light neutrino exchanges, are complicated and hard (in literatures to treat them even the Lorentz covariance is lost~\cite{Tomoda:1990rs}). In this work to calculate NMEs and PSF, we try to make an additional approximation on the contributions relevant the factors $\frac{k\!\!\!/}{k^2}P_{L,R}$ for the light neutrino exchanges, which we call as `frozen approximation'. Under the approximation, the momenta of the two involved quarks inside the initial nucleus and two involved quarks inside the final nucleus (Fig.~\ref{FeynmanLRSM}) are assumed to be equal approximately:
\begin{equation}
p_1\simeq p_2\equiv \bar{p},\;\;k_1\simeq k_2\equiv \overline{k}.\label{frozen}
\end{equation}
With the `frozen approximation' and `the on-shell approximation' onto the momenta for the out legs as well, all contributions corresponding to Fig.~\ref{FeynmanLRSM} (b) can be well-deduced and the final results can be collected as
\begin{eqnarray}
&&\frac{2m_p}{G_F^2\cos\theta_C^2}\mathcal{L}_{\rm eff}^{\rm LR}(\nu)=\frac{2m_p m_e}{-k^2}[C_{3R}^{LL}(\nu)\mathcal{O}_{3R}^{LL}+C_{3L}^{LL}(\nu)\mathcal{O}_{3L}^{LL}+
C_{3L}^{RL}(\nu)\mathcal{O}_{3L}^{RL}+C_{3R}^{RL}(\nu)\mathcal{O}_{3R}^{RL}+\nonumber\\
&&\qquad\qquad\qquad\qquad\quad C_5^{LL}(\nu)\mathcal{O}_{5}^{LL}+C_5^{RR}(\nu)\mathcal{O}_{5}^{RR}+C_5^{LR}(\nu)\mathcal{O}_{5}^{LR}+
C_5^{RL}(\nu)\mathcal{O}_{5}^{RL}],
\label{eq18}
\end{eqnarray}
where
\begin{eqnarray}
&&\mathcal{O}_{5}^{XY}=4(\bar u \gamma_{\mu} P_X d)(\bar u P_Y d)\bar e \gamma_\mu\gamma^5 e^c,\nonumber\\
&&C_{3R}^{LL}(\nu)=\frac{1}{m_e}\cos^3\zeta U_{1i}(m_{\nu_i}\cos\zeta U_{1i}-m_e \sin\zeta T^*_{1i}),\nonumber\\
&&C_{3L}^{LL}(\nu)=-\cos^3\zeta\sin\zeta U_{1i} T^*_{1i},\nonumber\\
&&C_{3L}^{RL}(\nu)=C_{3R}^{RL}(\nu)=-\frac{1}{2}\cos^4\zeta U_{1i}T^*_{1i}(\frac{M_{W_1}}{M_{W_2}})^2,\nonumber\\
&&C_5^{RR}(\nu)=-C_5^{LL}(\nu)=\frac{m_u-m_d}{m_e}\cos^3\zeta\sin\zeta U_{1i} T^*_{1i}-\frac{m_d}{m_e}\cos^4\zeta U_{1i}T^*_{1i}(\frac{M_{W_1}}{M_{W_2}})^2,\nonumber\\
&&C_5^{LR}(\nu)=-C_5^{RL}(\nu)=\frac{m_d-m_u}{m_e}\cos^3\zeta\sin\zeta U_{1i} T^*_{1i}-\frac{m_u}{m_e}\cos^4\zeta U_{1i}T^*_{1i}(\frac{M_{W_1}}{M_{W_2}})^2.\label{eq49}
\end{eqnarray}
Then in LRSM the half-life of $0\nu2\beta$ decays can be written as~\cite{Deppisch:2020ztt}
\begin{eqnarray}
&&\frac{1}{T_{1/2}^{0\nu}}=G^{0\nu}|M^{0\nu}|^2\Big|\frac{m_{ee}^{\rm LR}}{m_e}\Big|^2,
\end{eqnarray}
where
\begin{eqnarray}
&&m_{ee}^{\rm LR}=m_e\Big\{\Big|[C_{3L}^{RR}(N)+C_{3L}^{LL}(N)]U_3^{XX}\frac{M_3^{XX}(N)}{M^{0\nu}}+
C_{3L}^{RL}(N)U_{(31)11}^{XY}\frac{M_3^{XY}(N)}{M^{0\nu}}\nonumber\\
&&\qquad\quad+C_{3L}^{LL}(\nu)+C_{3L}^{RL}(\nu)\frac{M_3^{XY}(\nu)}
{M^{0\nu}}\Big|^2+\Big|C_{3R}^{LL}(N)U_3^{XX}\frac{M_3^{XX}(N)}{M^{0\nu}}+
C_{3R}^{LL}(\nu)\Big|^2\nonumber\\
&&\qquad\quad+2\frac{G_{11-}^{(0)}}{G^{0\nu}}
\Big[[C_{3L}^{RR}(N)+C_{3L}^{LL}(N)]U_3^{XX}\frac{M_3^{XX}(N)}{M^{0\nu}}+C_{3L}^{RL}(N)U_{(31)11}^{XY}\frac{M_3^{XY}(N)}
{M^{0\nu}}\nonumber\\
&&\qquad\quad+C_{3L}^{LL}(\nu)+C_{3L}^{RL}(\nu)\frac{M_3^{XY}(\nu)}{M^{0\nu}}\Big]
\Big[C_{3R}^{LL}(N)U_3^{XX}\frac{M_3^{XX}(N)}{M^{0\nu}}+C_{3R}^{LL}(\nu)\Big]\Big\}^{1/2},\label{meeLR}
\end{eqnarray}
where $U_{(31)}^{XY}$ is the $2\times2$ RGE evolution matrix from $\mu\simeq M_{W_1}$ to $\mu\simeq 1.0\;$GeV (the numerical result of $U_{(31)}^{XY}$ can be found in Eq.~(\ref{48})), $C_{3Z}^{XY}(N,\nu)$ is the coefficient defined in Eq.~(\ref{eq49}), $G^{(0)}_{11-}=-0.28\times10^{-15}\;(-1.197\times10^{-15})\;{\rm years}^{-1}$~\cite{Deppisch:2020ztt} for $^{76}{\rm Ge}\;(^{136}{\rm Xe})$ is the PSF, and $M_3^{XY}(\nu)=4.24\pm0.68\;(2.17\pm0.35)$, $M_3^{XY}(N)=99.8\pm27.94\;(51.2\pm14.34)$~\cite{Barea:2015kwa,Deppisch:2020ztt} for $^{76}{\rm Ge}\;(^{136}{\rm Xe})$ are the NMEs corresponding to the exchange of light neutrinos, heavy neutral leptons respectively which are defined as
\begin{eqnarray}
&&M_3^{XY}(\nu)\equiv\langle{\mathcal O}_F^+|\frac{2m_p m_e}{-k^2}[4(\bar u \gamma_{\mu} P_X d)
(\bar u \gamma^{\mu} P_Y d)]|\mathcal O_I^+\rangle,\nonumber\\
&&M_3^{XY}(N)\equiv\langle{\mathcal O}_F^+|[4(\bar u \gamma_{\mu} P_X d)
(\bar u \gamma^{\mu} P_Y d)]|\mathcal O_I^+\rangle.
\end{eqnarray}
Similar to the case of B-LSSM, the factor $\frac{2m_p m_e}{-k^2}$ in Eq.~(\ref{eq18}) is absorbed into the `neutrino potential' which is used to compute the long range NME. In addition, we should note that in Eq.~(\ref{meeLR}) the terms of $C_5^{RR}(\nu), C_5^{LL}(\nu), C_5^{RL}(\nu), C_5^{LR}(\nu)$ in Eq.~(\ref{eq49}) do not make contributions to the $0\nu2\beta$ decays for the reason below. They make contributions to $0\nu2\beta$ decays in the form
\begin{eqnarray}
&&M_5^{XX}(\nu)[C_5^{RR}(\nu)+C_5^{LL}(\nu)]+M_5^{XY}(\nu)[C_5^{RL}(\nu)+C_5^{LR}(\nu)],\label{222}
\end{eqnarray}
where
\begin{eqnarray}
&&M_5^{XX}(\nu)\equiv\langle{\mathcal O}_F^+|\frac{2m_p m_e}{-k^2}[4(\bar u \gamma_{\mu} P_X d)
(\bar u P_X d)]|\mathcal O_I^+\rangle,\nonumber\\
&&M_5^{XY}(\nu)\equiv\langle{\mathcal O}_F^+|\frac{2m_p m_e}{-k^2}[4(\bar u \gamma_{\mu} P_X d)
(\bar u P_Y d)]|\mathcal O_I^+\rangle.
\end{eqnarray}
Since $C_5^{RR}(\nu)=-C_5^{LL}(\nu)$, $C_5^{RL}(\nu)=-C_5^{LR}(\nu)$ as shown in Eq.~(\ref{eq49}), then Eq.~(\ref{222}) shows the contributions to the decays from the operators corresponding to the coefficients $C_5^{RR}(\nu)$, $C_5^{LL}(\nu)$, $C_5^{RL}(\nu)$, $C_5^{LR}(\nu)$ are cancelled completely.

\section{Numerical results\label{sec4}}

With the formulas obtained above, now in this section we do the numerical calculations and present the results on the $0\nu2\beta$ decays for the the nuclei $^{76}{\rm Ge}$ and $^{136}{\rm Xe}$ accordingly.
In our numerical calculations, the parameters are taken as the weak boson mass: $M_{W_L}=80.385$ GeV for B-LSSM and $M_{W_1}=80.385$ GeV for LRSM,
$m_b=4.65$ GeV for b-quark mass, $m_c=1.275$ GeV for c-quark mass, $\alpha_{em}(m_Z)=1/128.9$ for the coupling of the electromagnetic
interaction, $\alpha_{s}(m_Z)=0.118$ for the coupling of the strong interaction. The constraints from available experimental data, such as the most stringent upper limit on the sum of neutrino masses by PLANK~\cite{Esteban:2018azc}
$\sum_i m_{\nu_i}<0.12\;{\rm eV}$; the neutrino mass-squared differences obtained
via analyzing the solar and atmospheric neutrino oscillation data at $3\sigma$ deviations~\cite{Esteban:2018azc}
\begin{eqnarray}
&&\Delta m_{12}^2\equiv m_{\nu_2}^2-m_{\nu_1}^2=(7.4\pm0.61)\times 10^{-5}\;{\rm eV}^2,\nonumber\\
&&\left\{\begin{array}{l}
\Delta m_{13}^2\equiv m_{\nu_3}^2-m_{\nu_1}^2\approx(2.526\pm0.1)\times 10^{-3}\;{\rm eV}^2\qquad\qquad({\rm NH})\\
\Delta m_{32}^2\equiv m_{\nu_2}^2-m_{\nu_3}^2\approx(2.508\pm0.1)\times 10^{-3}\;{\rm eV}^2\qquad\qquad({\rm IH}),\\
\end{array}\right. \label{eq52}
\end{eqnarray}
etc, are well-considered. Owing to the fact that the hierarchy of neutrino masses has not
been fixed yet, we take the two possibilities below to carry out the analyses i.e.
the normal hierarchy (NH) $m_{\nu_1}<m_{\nu_2}<m_{\nu_3}$ and the inverse hierarchy
(IH) $m_{\nu_3}<m_{\nu_1}<m_{\nu_2}$. Moreover, we use the matrix $U$, the upper-left
sub-matrix of the whole matrix $U_\nu$ in Eq.~(\ref{eq6}), being the Pontecorvo-Maki-Nakagawa-Sakata (PMNS)
mixing matrix~\cite{Tanabashi:2018oca} to describe the mixing of the light neutrinos. Then we have
\begin{eqnarray}
&&(U_{1i})^2 m_{\nu i}=c_{12}^2c_{13}^2e^{i\alpha}m_{\nu_1}+c_{13}^2s_{12}^2e^{i\beta}m_{\nu_2}+s_{13}^2m_{\nu_3},
\end{eqnarray}
where $\alpha,\beta$ are Majorana CP violating phases,
$c_{12,13}\equiv\cos\theta_{12,13},s_{12,13}\equiv\sin\theta_{12,13}$ are the neutrino oscillation
parameters and at $3\sigma$ error level they can be written as
\begin{eqnarray}
&&s_{12}^2=0.3125\pm0.0375,\;s_{13}^2=0.022405\pm0.001965.
\end{eqnarray}

The direct searching for the right-handed gauge boson sets the lower bound for the mass of
$W_2$ boson as $M_{W_2}\gtrsim4.8\;{\rm TeV}$~\cite{Sirunyan:2017yrk,Sirunyan:2018pom,Aaboud:2018spl,Aaboud:2019wfg},
and the $W_R$, $W_L$ mixing angle as $\zeta\lesssim7.7\times10^{-4}$~\cite{Dev:2014xea}. Finally we have
$x\equiv v_2/v_1>0.02$~\cite{Bertolini:2014sua} required by the CP-violation data for the $K$ and
$B$ mesons, where $v_1$, $v_2$ are the VEVs of the two Higgs particles in the LRSM.

Eq.~(\ref{meeBL}) and Eq.~(\ref{meeLR}) shows $m_{ee}^{LR}$ depends the NME and phase space factor of the chosen nuclei. However these factors appears in the expression of $m_{ee}^{LR}$ in the form of some ratios, and these ratios for $^{76}{\rm Ge}$ are equal to those for $^{136}{\rm Xe}$ roughly, hence we can use the same expression of $m_{ee}^{LR}$ for $^{76}{\rm Ge}$, $^{136}{\rm Xe}$ in the analyses later on.

\subsection{The numerical results for the B-LSSM\label{sec4A}}

\begin{figure}
\setlength{\unitlength}{1mm}
\centering
\includegraphics[width=3in]{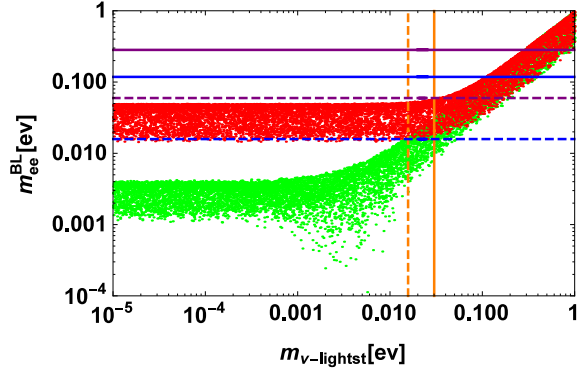}
\vspace{0cm}
\caption{$m_{ee}^{\rm BL}$ versus $m_{\nu-{\rm lightest}}$ under scanning the neutrino mass-squared differences Eq.~(\ref{eq52}) at $3\sigma$ error level and the parameter regions Eq.~(\ref{BLScaneq}). The green (red) points denote the NH (IH) results, the blue (purple) solid line denotes the experimental constraints from the $0\nu2\beta$ decay half-life of $^{76}{\rm Ge}$ ($^{136}{\rm Xe}$), the blue (purple) dashed line denotes the experimental ability of $^{76}{\rm Ge}$ ($^{136}{\rm Xe}$) for the next generation of experiments, the orange solid (dashed) line denotes the constraints from PLANK 2018 for the case of NH (IH).}
\label{BL}
\end{figure}
According to the above analysis, in B-LSSM the contributions from the heavy neutral leptons
shown in Eq.~(\ref{eq45}) are highly suppressed ($S_{1i}\approx 10^{-7}$)
for the TeV-scale heavy neutral leptons, hence the dominant contributions to
the $0\nu2\beta$ decays come from the light neutrinos. Then with the neutrino mixing parameters
\begin{eqnarray}
&&s_{12}^2=(0.275\sim 0.35),\;\;s_{13}^2=(0.02044\sim 0.02437),\;\;\alpha=(0\sim 2\pi),\;\;\beta=(0\sim 2\pi).\label{BLScaneq}
\end{eqnarray}
and the neutrino mass-squared differences as those in Eq.~(\ref{eq52}) at $3\sigma$ error level, $m_{ee}^{\rm BL}$ versus
$m_{\nu-{\rm lightest}}$ (the lightest neutrino mass) for the B-LSSM is plotted in Fig.~\ref{BL}, where the
green (red) points denote the NH (IH) results, the blue (purple) solid line denotes the constraints from the
lower $0\nu2\beta$ decay half-life bound of $^{76}{\rm Ge}$ ($^{136}{\rm Xe}$), the blue (purple) dashed line denotes the experimental
ability of $^{76}{\rm Ge}$ ($^{136}{\rm Xe}$) for the next generation of experiments, the orange
solid (dashed) line denotes the constraints from PLANK 2018 for NH (IH) neutrino masses
(the meaning for the red points, green points, blue lines, purple lines, orange lines is also
adopted accordingly in the figures later on). In the figure Fig.~\ref{BL}, $m_{ee}^{\rm BL}$ is well-below the experimental
upper bounds in the cases of NH and IH, and there is not tighter restriction on the $m_{\nu-{\rm lightest}}$
than that offered by PLANK. Additionally, the blue dashed line shows that there is certain opportunity
to observe the $0\nu2\beta$ decays in the next generation of experiments, whereas if any of the decays
is not observed in the next generation of experiments, then the IH neutrino masses will be excluded
completely by the blue dashed line in Fig.~\ref{BL}.

\subsection{The numerical results for the LRSM\label{sec4B}}

In the LRSM owing to the bosons $W_L$, $W_R$ and their mixing, the situation is much more complicated
than that in the B-LSSM, and both of the light neutrinos and heavy neutral leptons make
substantial contributions to the $0\nu2\beta$ decays. For TeV-scale heavy neutral leptons in the LRSM,
the consequences of the type II seesaw dominance are similar to those of the type I seesaw dominance,
while the consequences are very different from the ones of type I$+$II seesaw dominance, because the light-heavy
neutral lepton mixing is not much suppressed in this case. Hence we do the numerical computations for
the type I and type I$+$II seesaw dominance cases in our analyses. For simplicity and not
losing general features, we assume that there is only one of heavy neutral leptons to make substantial
contributions in the model. It indicates that there are no off-diagonal elements in the matrix $M_R$
in Eq.~(\ref{eq11}), i.e. we have $M_R\simeq \hat M_N={\rm diag}(M_{N_1},M_{N_2},M_{N_3})$. Then under the dominance of either the type I seesaw or the type I+II seesaw mass mechanisms, we compute $m_{ee}^{\rm LR}$ numerically in tern.

\subsubsection{The results under type I seesaw dominance\label{sec4B1}}

\begin{figure}
\setlength{\unitlength}{1mm}
\centering
\includegraphics[width=3in]{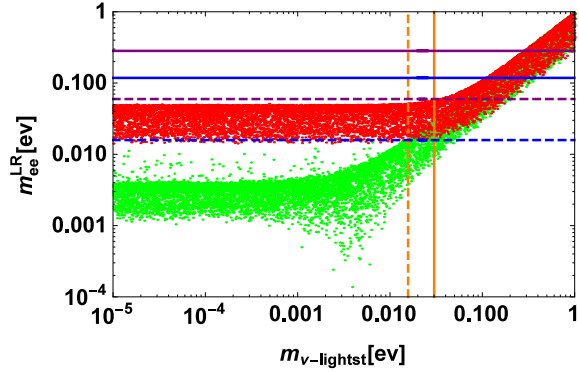}
\vspace{0cm}
\caption{$m_{ee}^{\rm LR}$ versus $m_{\nu-{\rm lightest}}$ with scanning the parameter space in Eqs.~(\ref{eq52},~\ref{BLScaneq},~\ref{scaneqLR}).}
\label{LRI}
\end{figure}

Under the type I seesaw dominance and due to the tiny neutrino masses, the sub-matrix $T$
in Eq. (\ref{eq6}) has $T_{1j}\ll 1\;(j=1,2,3)$. It indicates that the contributions from
the terms proportional to $T_{1i}^*$ as shown in Eq.~(\ref{eq49}) are highly suppressed.
Then scanning the parameter space in Eqs.~(\ref{eq52},~\ref{BLScaneq}) and in the
parameter space
\begin{eqnarray}
x=(0.02\sim 0.5),\;\;M_{W_2}=(4.8\sim 10.0)\;{\rm TeV},\;\;M_{N_1}=(0.10\sim 3.0)\;{\rm TeV},\label{scaneqLR}
\end{eqnarray}
we plot $m_{ee}^{\rm LR}$ versus $m_{\nu-{\rm lightest}}$ in Fig.~\ref{LRI}.
Comparing Fig.~\ref{LRI} with Fig.~\ref{BL}, the red points show that the range of $m_{ee}^{\rm LR}$
is similar to the range of $m_{ee}^{\rm BL}$ in the case of IH neutrino masses, but in the case of
NH neutrino masses there are points with $m_{ee}^{\rm LR}>m_{ee}^{\rm BL}$. For IH neutrino masses,
the contributions to $m_{ee}^{\rm LR}$ are dominated by the terms proportional to light neutrino
masses, hence $m_{ee}^{\rm LR}$ depends on $x,M_{W_2},M_{N_1}$ negligibly, that leads to the the
range of reds points is similar to the results of $m_{ee}^{\rm BL}$. Additionally, $m_{ee}^{\rm LR}>m_{ee}^{\rm BL}$
shown as green points indicates the contributions to $m_{ee}^{\rm LR}$ can be dominated by heavy neutral
leptons for appropriate values of $x,M_{W_2},M_{N_1}$.

\begin{figure}
\setlength{\unitlength}{1mm}
\centering
\includegraphics[width=2.5in]{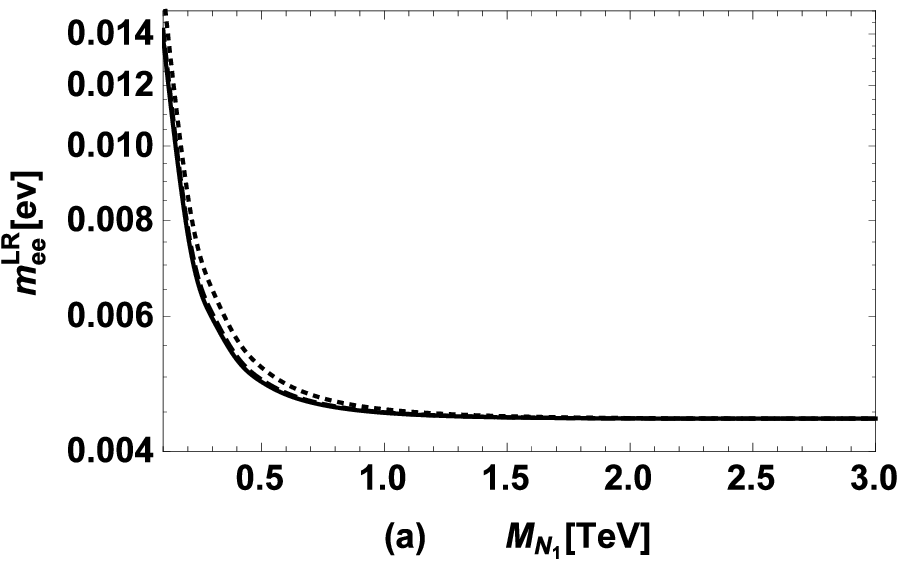}
\vspace{0.5cm}
\includegraphics[width=2.5in]{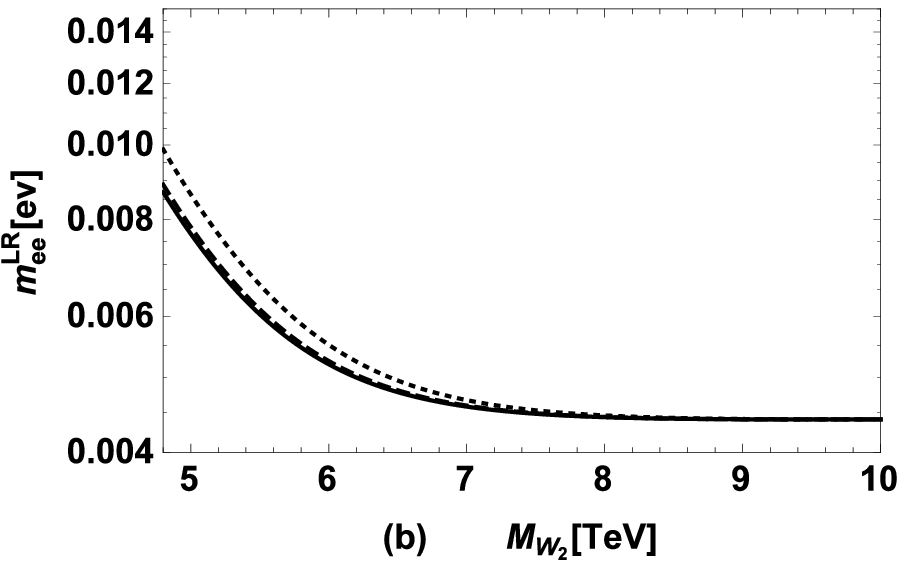}
\vspace{0cm}
\caption[]{With $m_{\nu_1}=0.001\;{\rm eV}$ for NH neutrino masses, (a): $m_{ee}^{\rm LR}$ versus $M_{N_1}$ for $M_{W_2}=5.0\;{\rm TeV}$, (b): $m_{ee}^{\rm LR}$ versus $M_{W_2}$ for $M_{N_1}=0.2\;{\rm TeV}$. The solid, dashed, dotted lines denote the obtained $m_{ee}^{\rm LR}$ for $x=0.10,\,0.25,\,0.40$ respectively.}
\label{I-Mnu}
\end{figure}

In order to see the effects of $x,M_{W_2},M_{N_1}$ clearly, we take $m_{\nu_1}=0.001\;{\rm eV}$ for
the NH neutrino masses, $s_{12},s_{13},\Delta m_{12}^2,\Delta m_{13}^2$ at the corresponding center
values and the CP violation phases $\alpha=\beta=0$. Then taking $M_{W_2}=5.0\;{\rm TeV}$, we plot
$m_{ee}^{\rm LR}$ versus $M_{N_1}$ in Fig.~\ref{I-Mnu} (a), where the solid, dashed, dotted lines
denote the results for $x=0.10,\,0.25,\,0.40$ respectively. Similarly, $m_{ee}^{\rm LR}$ versus $M_{W_2}$
for $M_{N_1}=0.20\;{\rm TeV}$ is plotted in Fig.~\ref{I-Mnu} (b).
From Fig.~\ref{I-Mnu}, one may see the fact that the obtained $m_{ee}^{\rm LR}$ decreases with
the increasing of $M_{N_1}$, $M_{W_2}$, then $m_{ee}^{\rm LR}$ approaches to a constant when
$M_{N_1}$ or $M_{W_2}$ becomes large. According to the definition of $W_L$-$W_R$
mixing parameter $\zeta$, the coefficient $C_{3R}^{RL}(N)$ in Eq.~(\ref{eq47}) increases with
increasing $x\equiv \frac{v_2}{v_1}$, which leads to that $m_{ee}^{\rm LR}$ increases with $x$ increasing as shown in the figures. Additionally one may see that $m_{ee}^{\rm LR}$ depends on the values of $x,\;M_{W_2},\;M_{N_1}$
mildly when $M_{N_1}$ or $M_{W_2}$ is large. It is because that the contributions from heavy
neutral lepton(s) are highly suppressed when its mass or right handed boson mass becomes large,
thus the contributions from the light neutrinos become dominant and proportional merely to the
light neutrino masses in this case.

\subsubsection{The results under the type I+II seesaw dominance\label{sec4B2}}

\begin{figure}
\setlength{\unitlength}{1mm}
\centering
\includegraphics[width=3in]{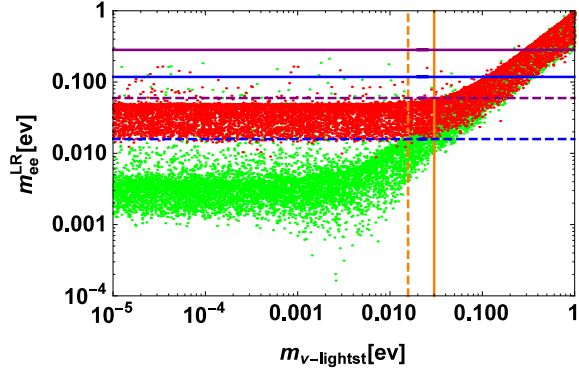}
\vspace{0cm}
\caption{$m_{ee}^{\rm LR}$ versus $m_{\nu-{\rm lightest}}$ with scanning the parameter space in Eqs.~(\ref{eq52},~\ref{BLScaneq},~\ref{scaneqLRII}).}
\label{LRII+I}
\end{figure}

As pointed out above, the mixing parameters of the light neutrinos and the heavy neutral leptons are not tiny under the type I$+$II seesaw dominance, hence the Dirac mass matrix $M_D$ can also affect the numerical results via the mixing of the light neutrinos and the heavy neutral leptons. For simplicity and not losing general feature, we assume that the mass matrix $M_D$ in Eq.~(\ref{eq11}) is diagonal as $M_D={\rm diag}(M_{D11},M_{D22},M_{D33})$. Taking the assumption that only one generation of heavy neutral leptons makes substantial contributions, i.e. $M_{D22}$ and $M_{D33}$ do not affect the results. Scanning the ranges determined by Eqs.~(\ref{eq52},~\ref{BLScaneq}) and the parameter space
\begin{eqnarray}
&&x=(0.02\sim 0.5),\,M_{W_2}=(4.8\sim 10)\;{\rm TeV},M_{N_1}=(0.10\sim 3.0)\;{\rm TeV},\nonumber\\
&&M_{D11}=(0.10\sim 100)\;{\rm MeV},\label{scaneqLRII}
\end{eqnarray}
$m_{ee}^{\rm LR}$ versus $m_{\nu-{\rm lightest}}$ as the results is plotted in Fig.~\ref{LRII+I}.
Comparing Fig.~\ref{LRII+I} with Fig.~\ref{LRI} (also Fig.~\ref{BL} ), a lager $m_{ee}^{\rm LR}$
can be reached in the case of type I+II seesaw dominance. Because the contributions are dominated
by the terms proportional to $S_{1i}$ or $T_{1i}^*$  when $M_{D11}$ is large (shown in Eqs.~(\ref{eq47},~\ref{eq49})), and lager $m_{ee}^{\rm LR}$ is reached in this case. As indicated in Fig.~\ref{LRII+I} there are some points which are smaller than the points in Fig.~\ref{LRI} both for NH and IH neutrino masses, that is due to the existence of the cancellation effect, i.e. the contributions from the
terms proportional to $M_{D11}$ can be cancelled some amount of the contributions from the other terms. The parameter $M_{D11}$ represents the strength of light-heavy neutral lepton mixing, that we can introduce a familiar parameter $S_e^2\equiv\sum_{i=1}^3|S_{1i}|^2$ for the following analysis to describe the light-heavy neutral lepton mixing.

\begin{figure}
\setlength{\unitlength}{1mm}
\centering
\includegraphics[width=2.5in]{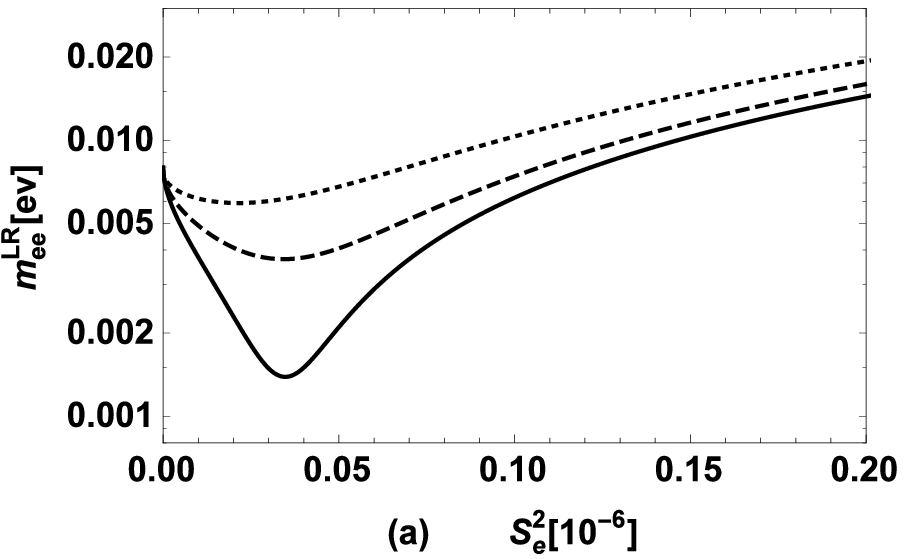}
\vspace{0.5cm}
\includegraphics[width=2.5in]{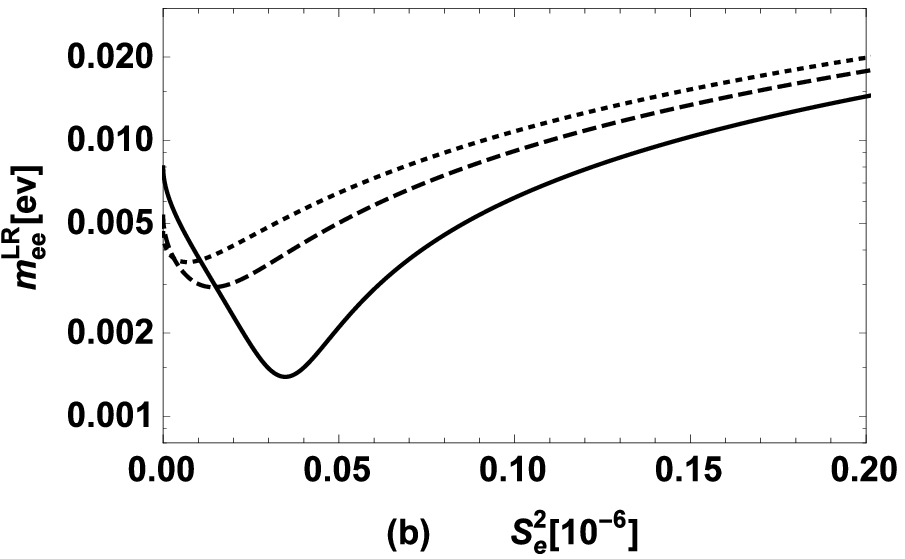}
\vspace{0cm}
\caption[]{$m_{ee}^{\rm LR}$ versus $S_e^2$ with $M_{N_1}=0.2\;{\rm TeV}$, $m_{\nu_1}=0.001\;{\rm eV}$ for the NH neutrino masses, $s_{12}$, $s_{13}$, $\Delta m_{12}^2$, $\Delta m_{13}^2$ at the corresponding center values and the CP violation phases $\alpha=\beta=0$. (a): the results for $M_{W_2}=5.0\;{\rm TeV}$, and the solid, dashed, dotted lines denote the results for $x=0.02,\,0.06,\,0.10$ respectively. (b): the results for $x=0.02$, and the solid, dashed, dotted lines denote the results for $M_{W_2}=5.0,\,6.0,\,7.0\;{\rm TeV}$ respectively.}
\label{II+I-zeta}
\end{figure}

In order to show the effects of light-heavy mixing parameter $S_e^2$ and the cancellation effect well, we take $M_{N_1}=0.2\;{\rm TeV}$, $m_{\nu_1}=0.001\;{\rm eV}$ for the NH neutrino masses, $s_{12}^2,\,s_{13}^2,\,\Delta m_{12}^2,\,\Delta m_{13}^2$ with the center values and
the CP violation phases $\alpha=\beta=0$ accordingly. The effects with the parameters being fixed above are presented a similar behavior as the case of type I seesaw dominance, hence we do not repeat the study of the effects on these parameters. With $M_{W_2}=5.0\;{\rm TeV}$, in Fig.~\ref{II+I-zeta} (a) we plot $m_{ee}^{\rm LR}$ versus $S_e^2$, and the solid, dashed, dotted lines denote the results for $x=0.02,\,0.06,\,0.1$
respectively. Similarly, with $x=0.02$, $m_{ee}^{\rm LR}$ versus $S_e^2$ is plotted
in Fig.~\ref{II+I-zeta} (b), where the solid, dashed, dotted lines denote the results for
$M_{W_2}=5.0,\,6.0,\,7.0\;{\rm TeV}$ respectively.

From Fig.~\ref{II+I-zeta}(a, b) one may see the fact that with the mixing parameter $S_e^2$ increasing, $m_{ee}^{\rm LR}$
decreases to a minimum value and then increases. In the case of the type I+II seesaw dominance, i.e.
$S_{1i}$ and $T_{1i}$ are not too small, the contributions from the light neutrinos are dominated over those terms
which do not depend on the neutrino mass $m_{\nu_i}$ in Eq. (\ref{eq49}). When $S_e^2$ is small,
$C_{3L}^{RR}(N)$ in Eq.~(\ref{eq47}) plays a dominant role, and as $S_e^2$ increasing, the contributions
from $C_{3L/R}^{LL}(\nu)$ in Eq.~(\ref{eq49}) become larger. The minimum values for $m_{ee}^{\rm LR}$ as shown in Fig.~\ref{II+I-zeta} is due to the opposite signs for $M_3^{XX}(N)$ and $M_3^{XY}(\nu)$, i.e. it is caused by cancellation of the contributions from $C_{3L}^{RR}(N)$ and $C_{3L/R}^{RL}(\nu)$. From the figures the fact can be seen clearly that when $S_e^2$ is increasing, the contributions from $C_{3L/R}^{RL}(\nu)$ become dominant which leads to the increasing of $m_{ee}^{\rm LR}$ as long as $S_e^2$ becomes large enough. The results indeed show that with $S_e^2$
varying, the cancellation takes place in a proper manner. When the cancellation takes place, the value
of $S_e^2$ depends on $M_{W_2}$ and $x$ explicitly.

\begin{figure}
\setlength{\unitlength}{1mm}
\centering
\includegraphics[width=2.5in]{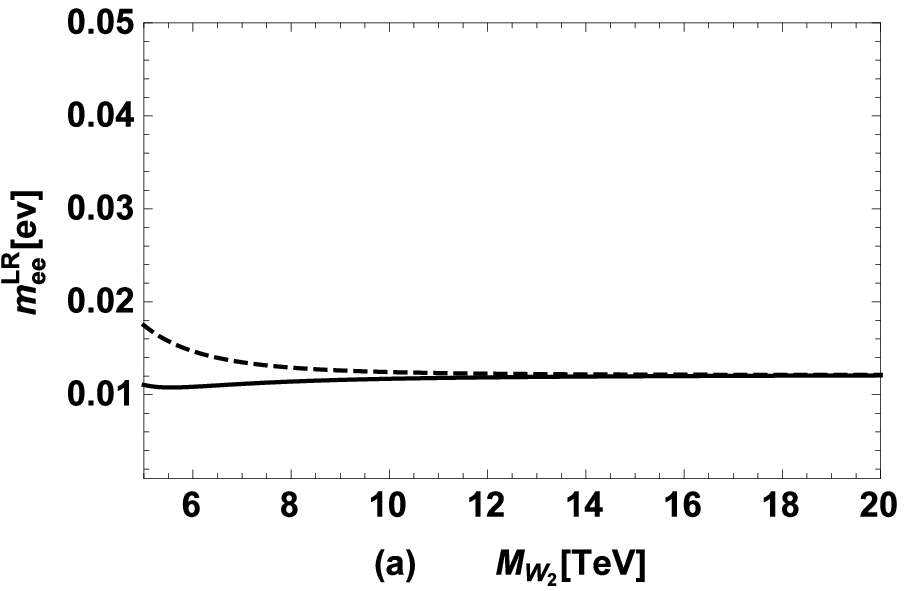}
\vspace{0.5cm}
\includegraphics[width=2.5in]{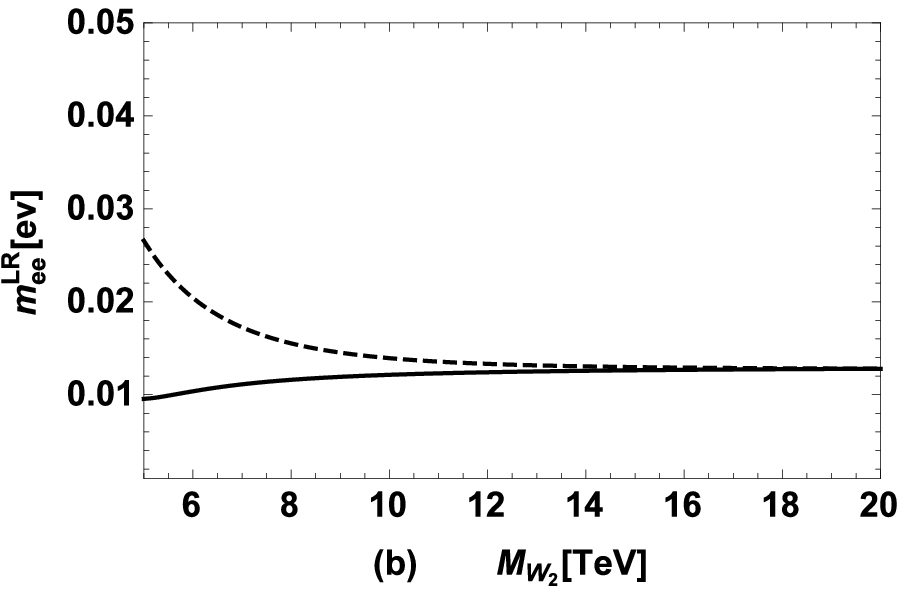}
\vspace{0cm}
\caption[]{(a): $m_{ee}^{\rm LR}$ versus $M_{W_2}$ for $S_e^2=0.25\times10^{-8}$. (b): $m_{ee}^{\rm LR}$ versus $M_{W_2}$ for $S_e^2=10^{-8}$. Both are with $M_{N_1}=0.2\;{\rm TeV}$, $m_{\nu_1}=0.01\;{\rm eV}$ for the NH neutrino masses.  The solid, dashed lines denote the results obtained under the new approximation in this work and the traditions method shown in Ref.~\cite{Dev:2014xea} respectively.}
\label{com}
\end{figure}

To compare the results obtained under the fresh approximation in this work with the ones obtained by traditional way, we take the results from Ref.~\cite{Dev:2014xea} and present $m_{ee}^{\rm LR}$ versus $M_{W_2}$ in Fig.~\ref{com} (a), (b) for $s_e^2=0.25\times10^{-8},\;10^{-8}$ respectively, where the solid, dashed lines denote the results obtained under the fresh approximation in this work and those by the traditions way from in Ref.~\cite{Dev:2014xea} respectively.
In Fig.~\ref{com}, we take $M_{N_1}=0.2\;{\rm TeV}$, $m_{\nu_1}=0.010\;{\rm eV}$, $x=0.020$, $s_{12},\,s_{13},\,\Delta m_{12}^2,\,\Delta m_{13}^2$ at the corresponding center values and the CP violation phases $\alpha=\beta=0$. Note that in the figures the results obtained in this work coincide well with the traditional ones when $S_e^2$ is small or $M_{W_2}$ is large in the chosen parameter space. In addition, the results obtained in this work, being approximate ones but having the interference effects considered better, are smaller than the ones obtained in traditional way when $W_2$ is not heavy enough, and the reduction factor depends on the parameters in LRSM completely and also comes from the uncertainties of NMEs partly. Considering the difficulties and the uncertainties etc in computing NMEs and the interference effects among the various contributions by the way in literatures and the approximation approach in this work, from Fig.\ref{com}, it seems that the results with very heavy $W_2$ ($M_{W_2} \geq 12$TeV) or small $S_e^2$ $(\lesssim 10^{-9})$ may be convinced more as the results approach to coinciding with each other.

\section{Summary and Discussions\label{sec5}}

In the paper we take B-L supersymmetric standard model (B-LSSM) and TeV scale left-right symmetric model (LRSM) as two representations of two kinds of new physics models to study the nuclear neutrinoless double beta decays ($0\nu2\beta$). As stated in Introduction, the calculations, on which are focused the lights, are those about the factor on the 'core' factor of the decays: evaluating the process $d+d\to u+u+e+e$ by considering the effective Lagrangian containing the operators with the Wilson coefficients or considering the relevant amplitude etc. Whereas here the estimations of the other necessary `factors' for the decays, i.e. to evaluate relevant nuclear matrix element (NME) and phase space factor (PSF) etc, that does not relate to the specific models, are treated by following literatures.

In the B-LSSM, all of the calculations can be well deduced and the results are dominated by the neutrino mass terms. However, in the LRSM owing to the existence of right-handed gauge boson $W_R$, the calculations are complicated, and the interference effects are hard and not well to be estimated. In this work, a new approximation, i.e. the momenta of the two involved quarks inside the initial nucleus and inside the final nucleus are assumed to be equal approximately, is made, then all contributions in the LRSM can be well reduced and summed up all the contributions. Then the calculations in LRSM are simplified and the interference effects can be calculated comparatively easily. To see the consequences of the approximation, we compare the results obtained in this work with the ones obtained by the traditional method numerically, and the results coincide with each other well when light-heavy neutral lepton mixing parameter $S_e^2$ is small or $M_{W_2}$ is large. For the effective dimension-9 contact interactions in these two models, the contributions from heavy neutral lepton exchange, the QCD corrections from the energy scale $\mu\simeq M_{W_L}$ (or $M_{W_1}$) to the energy scale $\mu\simeq 1.0\;$GeV to all dimension-9 operators in the effective Lagrangian which is responsible for the $0\nu2\beta$ decays are calculated by the RGE method, and all the QCD corrections including the contributions from light neutrnos in the energy scale region $\mu \simeq 1.0\;$GeV to $\mu \simeq 0.10\;$GeV, being of non-perturbative QCD, are taken into account alternatively by inputting in the experimental measurements for the relevant current matrix elements of nucleons, which emerge at the effective Lagrangian at $\mu\simeq 0.10\;$GeV.

With necessary input parameters allowed by experimental data, the theoretical predictions on $0\nu2\beta$ decay half life $^{76}{\rm Ge}$ and $^{136}{\rm Xe}$ is obtained in these two models. In the B-LSSM, the contributions from heavy neutral leptons are highly suppressed by the tiny light-heavy neutral lepton mixing parameters for TeV-scale heavy neutral leptons. Hence $m_{ee}^{\rm BL}$ depends on the light neutrino masses mainly, and the numerical results show that the $0\nu2\beta$ decays may be observed with quite great opportunity in the near future.
Whereas if the decays are not observed in the next generation of $0\nu2\beta$ experiments,
the IH neutrino masses are excluded completely by the lower bound on $T_{1/2}^{0\nu}(^{76}{\rm Ge})$ (as shown in Fig.~\ref{BL}).

In the LRSM, the situation is much more complicated than that in the B-LSSM, so we have learnt a lot of experiences in study the $0\nu2\beta$ decays for the model. As for the type I seesaw dominance, the contributions from the terms proportional to $T_{1i}^*$ (see Eq.~(\ref{eq49})) are highly suppressed by the tiny neutrino masses.
The numerical results of the contributions from light neutrinos are similar to the ones
for B-LSSM, but the heavy neutral leptons can make comparatively large contributions
through the right-handed current when $M_{N_i}$ ($\lesssim0.5\;{\rm TeV}$)
and $M_{W_2}$ ($\lesssim7\;{\rm TeV}$) both are not too heavy. For the type I+II seesaw dominance,
the terms which do not depend on the neutrino mass $m_{\nu_i}$ in Eq.~(\ref{eq49}) play the
dominant roles. In this case, the contributions from $C_{3L/R}^{RL}(L)$ can be cancelled
some by the contributions from $C_{3L}^{RR}(H)$ when the light-heavy mixing parameter $S_e^2$
is appropriate, because the signs of the corresponding NMEs are
opposite. Moreover the effects of the cancellation are affected by $S_e^2$, $x$
and the right-handed $W$-boson mass $M_{W_2}$ etc in a complicated way. In addition,
Fig.~\ref{LRII+I} shows that the points either in green (NH) or in red (IH) spread
out a lot, and there are many `exotic points', where cannot be reached for the
cases of the B-LSSM and type I seesaw dominance LRSM. Thus the characteristic feature
on the distribution of $m_{ee}^{\rm LR}$ versus $m_{\nu-{\rm lightest}}$ may help to
realize whether the decays are caused by the LRSM in the type I+II seesaw dominance
or not, particularly when the $0\nu2\beta$ decays are observed and the points for
$m_{ee}^{\rm LR}$ versus $m_{\nu-{\rm lightest}}$ just fall on the exotic points.

Finally, according to the numerical results of the present comparative studies on the $0\nu2\beta$ decays for the two
typical models LRSM and B-LSSM, it may be concluded that the room of LRSM type models for the foreseeable future decay
experiments is greater than that of B-LSSM type models, and having the right handed gauge bosons the feature of the
LRSM type models is more complicated than that of the B-LSSM type models.


\vspace{5mm}

\noindent {\bf\Large Acknowledgments:} This work was supported in part by the National Natural Science
Foundation of China (NNSFC) under Grants No. 11821505£¬ No. 12047503, No. 12075301
and No. 11705045. It was also supported in part by the Key Research Program
of Frontier Sciences, CAS, Grant No. QYZDY-SSW-SYS006. The authors (J.-L. Yang and C.-H. Chang) would
like to thank Prof. J.-H. Yu (ITP, CAS) for helpful discussions and suggestion.


\begin{appendix}
\section{QCD corrections to the relevant dimension-9 oprators\label{appA}}

The essential process $d+d\to u+u+e+e$ for the $0\nu2\beta$ decays involves six `fermion legs', so
the effective Lagrangian for the decays is composed generally by a set of independent dimension-9 operators
as follows~\cite{Pas:2000vn}:
\begin{eqnarray}
&&\mathcal{L}_{\rm eff}^{\rm DBD}=\frac{G_F^2\cos^2\theta_C}{2m_p}\sum_{X,Y,Z}
\Big[\sum_{i=1}^3C_{iZ}^{XY}(\mu)\cdot\mathcal{O}_{iZ}^{XY}(\mu)+
\sum_{j=4}^5C_j^{XY}(\mu)\cdot\mathcal{O}_j^{XY}(\mu)\Big],\label{1}
\end{eqnarray}
where $\theta_C$ is the Cabibbo angle, $\mu \simeq 0.1\;$GeV is the energy
scale where the decays take place, and the independent dimension-9 operators $\mathcal{O}_{iZ}^{XY}(\mu)$,
$\mathcal{O}_j^{XY}(\mu)$ are defined as:
\begin{eqnarray}
&&{\mathcal O}_{1Z}^{XY}(\mu)=4(\bar u P_X d)(\bar u P_Y d)j_Z,\nonumber\\
&&{\mathcal O}_{2Z}^{XX}(\mu)=4(\bar u \sigma_{\mu\nu} P_X d)
(\bar u \sigma^{\mu\nu} P_X d)j_Z,\nonumber\\
&&{\mathcal O}_{3Z}^{XY}(\mu)=4(\bar u \gamma_{\mu} P_X d)
(\bar u \gamma^{\mu} P_Y d)j_Z,\nonumber\\
&&{\mathcal O}_4^{XY}(\mu)=4(\bar u \gamma_{\mu} P_X d)
(\bar u \sigma^{\mu\nu} P_Y d)j_\nu,\nonumber\\
&&{\mathcal O}_5^{XY}(\mu)=4(\bar u \gamma_{\mu} P_X d)
(\bar u P_Y d)j^\mu,
\label{eq19}
\end{eqnarray}
where $X,Y,Z=L,R$\,; $P_{R/L}=(1\pm\gamma^5)/2$, the leptonic currents are defined as
\begin{eqnarray}
&&j_{R/L}=\bar e (1\pm\gamma^5) e^c,\qquad j_\mu=\bar e \gamma_\mu\gamma^5 e^c.\label{3}
\end{eqnarray}

In the region from $M_{W_L}$ to $1.0\;$GeV for the energy scale $\mu$, the pQCD is
applicable, so the QCD corrections to the effective Lagrangian for the $0\nu2\beta$ decays can be carried
out by renormalization group equation method~\cite{Buchalla:1995vs,Buras:1998raa}.
The corresponding QCD corrections for the $0\nu2\beta$ decays were also calculated
in Ref.~\cite{Gonzalez:2015ady}, so here we describe how the corrections are
calculated briefly. In RGE method the renormalized operator matrix elements $<\mathcal{O}_i>^{\rm (R)}$ ($\mathcal{O}_i$ is defined in Eq.~(\ref{eq19})) for pQCD relate to their bare ones up to one-loop level generally as the following form:
\begin{eqnarray}
&&<\mathcal{O}_i>^{\rm (R)}=\Big[\delta_{ij}+\frac{\alpha_s}{4\pi}b_{ij}(\frac{1}{\varepsilon}+\ln\frac{\mu^2}{-p^2})\Big]
<\mathcal{O}_j>^{\rm bare},\label{bareQCD}
\end{eqnarray}
where $<\mathcal{O}_j>^{\rm bare}$ are the `bare operator matrix elements'.
The `renormalization' for quark fields $q$ and operator elements $<\mathcal{O}_j>$ are given by
\begin{eqnarray}
&&q^{\rm bare}=Z_q^{1/2}q^{\rm R} \;\;(q=u,d),\;\;\;<\mathcal{O}_j>^{\rm bare}=Z_q^2Z_{ij}^{-1}<\mathcal{O}_j>^{\rm R},\label{6}
\end{eqnarray}
where
\begin{eqnarray}
&&Z_q=1-C_F\frac{\alpha_s}{4\pi}\frac{1}{\varepsilon}+\mathcal{O}(\alpha_s^2),
\end{eqnarray}
and $C_F=(N^2-1)/(2N)$ is the $SU(N)$ color factor ($N=3$). The singularities
in Eq.~(\ref{bareQCD}) are required to be cancelled, then we have
\begin{eqnarray}
&&Z_{ij}Z_q^{-2}=\Big[\delta_{ij}+\frac{\alpha_s}{4\pi}b_{ij}\frac{1}{\varepsilon}\Big].
\end{eqnarray}
Then $Z_{ij}$ can be read out and written as
\begin{eqnarray} Z_{ij}=\delta_{ij}+\frac{\alpha_s}{4\pi}(b_{ij}-2C_F\delta_{ij})\frac{1}{\varepsilon}+\mathcal{O}(\alpha_s^2).
\label{Zij}
\end{eqnarray}

Considering the 4-quark-leg operators $\mathcal{O}(q^{\rm bare})$ in the effective
Lagrangian Eq.~(\ref{1}) are constructed by the bare quark field $q^{\rm bare}$,
and the corresponding coefficients $C^{\rm bare}$ are also bare. Then $q^{\rm bare}$,
$C^{\rm bare}$ relate to the renormalized ones as
\begin{eqnarray}
&&q^{\rm bare}=Z_q^{1/2}q^{\rm R},\;\;\;\;C_i^{\rm bare}=Z_{ij}^C C_j^{\rm R}.
\end{eqnarray}
Hence we have
\begin{eqnarray}
&&C_k^{\rm bare}\mathcal{O}_k(q^{\rm bare})=Z_q^2Z_{ij}^CC_j^{\rm R}\mathcal{O}_i(q^{\rm R}),
\end{eqnarray}
and the matrix elements for the QCD corrections are read as
\begin{eqnarray}
&&\quad Z_q^2Z_{ij}^CC_j^{\rm R}<\mathcal{O}_i(q^{\rm R})>^{\rm bare}=C_j^{\rm R}<\mathcal{O}_i(q^{\rm bare})>^{\rm R},\label{11}
\end{eqnarray}
Combining Eq.~(\ref{11}) and Eq.~(\ref{6}), we can obtain
\begin{eqnarray}
&&Z_{ij}^C=Z_{ij}^{-1}.
\end{eqnarray}
Due to the fact that the bare quantities $C_i^{\rm bare}$ do not depend on
the renormalization energy scale $\mu$, we have
\begin{eqnarray}
&&\quad\;\frac{\rm d}{{\rm d}\ln\mu}C_i^{\rm bare}=\frac{\rm d}{{\rm d}\ln\mu}Z_{ij}^{-1}C_j^{\rm R}=0,
\end{eqnarray}
which can be rewritten as
\begin{eqnarray}
&&\frac{{\rm d}\vec C^{\rm R}(\mu)}{{\rm d}\ln\mu}=\hat\gamma^{\rm T}\vec C^{\rm R}(\mu).\label{RGE}
\end{eqnarray}
Eq.~(\ref{RGE}) is the RGE accordingly for the Wilson coefficients,
where $\vec C=(C_1,\;C_2,\cdot\cdot\cdot)$ is written as a vector form.
Then the anomalous dimension matrix $\hat\gamma$ can be written as
\begin{eqnarray}
&&\hat\gamma=\frac{1}{\hat Z}\frac{\rm d}{{\rm d}\ln\mu}\hat Z.
\end{eqnarray}
Combining with the one-loop expression in the $\overline{\rm{MS}}$-scheme
\begin{eqnarray}
&&\hat\gamma(\alpha_s)=-2\alpha_s\frac{\partial\hat Z_1(\alpha_s)}{\partial\alpha_s},
\end{eqnarray}
where $\hat Z_1$ is the coefficient matrix of $1/\varepsilon$ in Eq.~(\ref{Zij}),
the `anomalous dimension matrix' to the leading order can be written as
\begin{eqnarray}
&&\gamma_{ij}(\alpha_s)=\frac{\alpha_s}{4\pi}\gamma_{ij},\;\;\;{\rm with}\;\;\;\gamma_{ij}=-2(b_{ij}-2C_F\delta_{ij}).\label{anomalousmatrix}
\end{eqnarray}
Then solving Eq.~(\ref{RGE}), the evolution of Wilson coefficients $\vec C^{\rm R}(\mu)$
can be expressed by $\vec C^{\rm R}(\Lambda)$ in terms of the $\mu$-evolution matrix $\hat U(\mu,\Lambda)$:
\begin{eqnarray}
&&\vec C^{\rm R}(\mu)=\hat U(\mu,\Lambda)\cdot\vec C^{\rm R}(\Lambda),\label{EC}
\end{eqnarray}
where precisely
\begin{eqnarray}
&&\hat U(\mu,\Lambda)=\hat V{\rm Diag}\Big\{\Big[\frac{\alpha_s(\Lambda)}{\alpha_s(\mu)}\Big]^{\gamma_{i}/(2\beta_0)}\Big\}\hat V^{-1}\label{U},
\end{eqnarray}
and
\begin{eqnarray}
&&{\rm Diag}\{\gamma_i\}=\hat V^{-1}\hat \gamma \hat V,\label{A19}
\end{eqnarray}
with $\hat\gamma$ is $\gamma_{ij}$ in matrix form. The running
coupling constant to one-loop level of QCD can be written as
\begin{eqnarray}
&&\alpha_s(\mu)=\frac{\alpha_s(\Lambda)}{1-\beta_0\frac{\alpha_s(\Lambda)}{2\pi}\ln(\frac{\Lambda}{\mu})},
\end{eqnarray}
with $\beta_0=(33-2f)/3$, and $f$ is the number of the active quark flavors
which is varied with the energy scale $\mu$, and only the quark $q^f$ with
mass $m_f$ smaller than the upper bound of the considered energy scale region
are `active'. Thus in the region from $\sim 1.0\;$GeV to $m_c \sim 1.3\;$GeV we
have $f=3$, in the region from $m_c$ to $m_b \sim 4.6\;$GeV we have $f=4$ and in
the region from $m_b$ to $M_W$ we have $f=5$. As a result, the required matrix
to describe the QCD RGE evolution from $\mu=M_W$ to $\mu\simeq 1.0\;$GeV is
\begin{eqnarray}
&&\hat U(\mu,\Lambda=M_W)=\hat U^{f=3}(\mu,\mu_c)\hat U^{f=4}(\mu_c,\mu_b)\hat U^{f=5}(\mu_b,M_W).\label{QCD18}
\end{eqnarray}

\begin{figure}
\setlength{\unitlength}{1mm}
\centering
\includegraphics[width=5in]{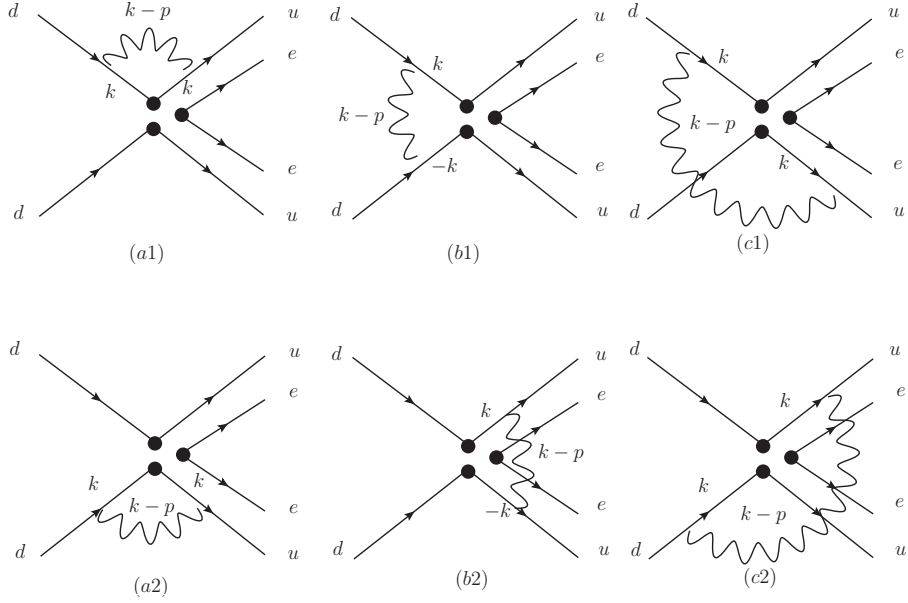}
\vspace{0cm}
\caption[]{One-loop QCD corrections to the dimension-9 operators for the $0\nu2\beta$ decays in the effective Lagragian.}
\label{QCDplot}
\end{figure}

To the leading order, the QCD corrections of the operators in Eq.~(\ref{eq19})
correspond to Fig.~\ref{QCDplot} as follows.
Calculating the diagrams respectively, the operator matrix elements corresponding
to those in Eq.~(\ref{bareQCD}) have the following structures
\begin{eqnarray}
&&{\rm Fig.~\ref{QCDplot}}(a1)\Rightarrow\mu^{4-D}\int\frac{{\rm d}^Dk}{(2\pi)^D}(\bar u\gamma_\alpha\frac{i}{k_\beta\gamma^\beta}\Gamma_i\frac{i}{k_\eta\gamma^\eta}\gamma^\alpha d)(\bar u\Gamma_j d)\cdot\frac{-i}{(k-p)^2}(ig_3)^2C_F\nonumber\\
&&\qquad\qquad\;\;=(\bar u\gamma_\alpha\gamma_\beta\Gamma_i\gamma^\beta\gamma^\alpha d)(\bar u\Gamma_j d)\cdot\frac{1}{4}C_F\frac{\alpha_s}{4\pi}(\frac{1}{\varepsilon}+\ln\frac{\mu^2}{-p^2}),\label{19}
\end{eqnarray}
\begin{eqnarray}
&&{\rm Fig.~\ref{QCDplot}}(a2)\Rightarrow\mu^{4-D}\int\frac{{\rm d}^Dk}{(2\pi)^D}(\bar u\Gamma_i d)(\bar u\gamma_\alpha\frac{i}{k_\beta\gamma^\beta}\Gamma_j\frac{i}{k_\eta\gamma^\eta}\gamma^\alpha d)\cdot\frac{-i}{(k-p)^2}(ig_3)^2C_F\nonumber\\
&&\qquad\qquad\;\;=(\bar u\Gamma_i d)(\bar u\gamma_\alpha\gamma_\beta\Gamma_j\gamma^\beta\gamma^\alpha d)\cdot\frac{1}{4}C_F\frac{\alpha_s}{4\pi}(\frac{1}{\varepsilon}+\ln\frac{\mu^2}{-p^2}),
\end{eqnarray}
\begin{eqnarray}
&&{\rm Fig.~\ref{QCDplot}}(b1)\Rightarrow\mu^{4-D}\int\frac{{\rm d}^Dk}{(2\pi)^D}(\bar u\Gamma_i\frac{i}{k_\eta\gamma^\eta}\gamma_\alpha T^a d)(\bar u\Gamma_j\frac{i}{-k_\sigma\gamma^\sigma}\gamma^\alpha T^a d)\cdot\frac{-i}{(k-p)^2}(ig_3)^2\nonumber\\
&&\qquad\qquad\;\;=-(\bar u\Gamma_i\gamma_\sigma\gamma_\alpha T^a d)(\bar u\Gamma_j \gamma^\sigma\gamma^\alpha T^a d)\cdot\frac{1}{4}\frac{\alpha_s}{4\pi}(\frac{1}{\varepsilon}+\ln\frac{\mu^2}{-p^2}),
\end{eqnarray}
\begin{eqnarray}
&&{\rm Fig.~\ref{QCDplot}}(b2)\Rightarrow\mu^{4-D}\int\frac{{\rm d}^Dk}{(2\pi)^D}(\bar u\gamma_\alpha\frac{i}{k_\eta\gamma^\eta}\Gamma_i T^a d)(\bar u\gamma^\alpha\frac{i}{-k_\sigma\gamma^\sigma}\Gamma_j T^a d)\cdot\frac{-i}{(k-p)^2}(ig_3)^2\nonumber\\
&&\qquad\qquad\;\;=-(\bar u\gamma_\alpha\gamma_\sigma\Gamma_i T^a d)(\bar u \gamma^\alpha\gamma^\sigma\Gamma_j T^a d)\cdot\frac{1}{4}\frac{\alpha_s}{4\pi}(\frac{1}{\varepsilon}+\ln\frac{\mu^2}{-p^2}),
\end{eqnarray}
\begin{eqnarray}
&&{\rm Fig.~\ref{QCDplot}}(c1)\Rightarrow\mu^{4-D}\int\frac{{\rm d}^Dk}{(2\pi)^D}(\bar u\Gamma_i\frac{i}{k_\eta\gamma^\eta}\gamma_\alpha T^a d)(\bar u\gamma^\alpha\frac{i}{k_\sigma\gamma^\sigma}\Gamma_j T^a d)\cdot\frac{-i}{(k-p)^2}(ig_3)^2\nonumber\\
&&\qquad\qquad\;\;=(\bar u\Gamma_i\gamma_\sigma\gamma_\alpha T^a d)(\bar u\gamma^\alpha\gamma^\sigma\Gamma_j T^a d)\cdot\frac{1}{4}\frac{\alpha_s}{4\pi}(\frac{1}{\varepsilon}+\ln\frac{\mu^2}{-p^2}),
\end{eqnarray}
\begin{eqnarray}
&&{\rm Fig.~\ref{QCDplot}}(c2)\Rightarrow\mu^{4-D}\int\frac{{\rm d}^Dk}{(2\pi)^D}(\bar u\gamma_\alpha\frac{i}{k_\eta\gamma^\eta}\Gamma_i T^a d)(\bar u\Gamma_j\frac{i}{k_\sigma\gamma^\sigma}\gamma^\alpha T^a d)\cdot\frac{-i}{(k-p)^2}(ig_3)^2\nonumber\\
&&\qquad\qquad\;\;=(\bar u\gamma_\alpha\gamma_\sigma\Gamma_i T^a d)(\bar u\Gamma_j\gamma^\sigma\gamma^\alpha T^a d)\cdot\frac{1}{4}\frac{\alpha_s}{4\pi}(\frac{1}{\varepsilon}+\ln\frac{\mu^2}{-p^2}),\label{24}
\end{eqnarray}
where $\Gamma_i$ are the Lorentz structures of the operators in Eq.~(\ref{eq19}),
and $T^a$ are the generators of $SU(N)$. Since the lepton sector is irrelevant with
the QCD corrections, so in the calculation the leptonic factor in the operators is
irrelevant. According to Eqs.~(\ref{bareQCD}$-$\ref{anomalousmatrix}), the anomalous
dimension matrix elements can be extracted from Eqs.~(\ref{19}$-$\ref{24}).

Summarizing the obtained anomalous dimension matrix elements for
$\mathcal{O}_{1Z}^{XY},\;\mathcal{O}_{2Z}^{XX},\;\mathcal{O}_{3Z}^{XY}$,
$\mathcal O_4^{XY}$ and $\mathcal O_5^{XY}$, we have
\begin{eqnarray}
&&\hat\gamma_{(12)}^{XX}=-2\left(\begin{array}{c}6C_F-3,\;\;-\frac{1}{2N}+\frac{1}{4}\\-12-\frac{24}{N},\;\;
-3-2C_F\end{array}\right),
\hat\gamma_{(31)}^{XY}=-2\left(\begin{array}{c}-\frac{3}{N},\;\;-6\\\;\;\;\;0,\;\;
\;\;\;6C_F\end{array}\right),\nonumber\\
&&\gamma_{(3)}^{XX}=-2\Big(\frac{3}{N}-3\Big),\nonumber\\
&&\hat\gamma_{(45)}^{XX}=-2\left(\begin{array}{c}-\frac{3}{2}-C_F,\;\;-\frac{3}{2}i-\frac{3i}{N}\\-\frac{i}{2}+\frac{i}{N},\;\;
\;\;3C_F-\frac{3}{2}\end{array}\right),
\hat\gamma_{(45)}^{XY}=-2\left(\begin{array}{c}-\frac{3}{2}-C_F,\;\;\frac{3}{2}i+\frac{3i}{N}\\ \frac{i}{2}-\frac{i}{N},\;\;
\;\;3C_F-\frac{3}{2}\end{array}\right).\label{eq50}
\end{eqnarray}
It is easy to realize that our results on the anomalous dimensions for the matrix elements $\mathcal O_{4,5}^{XY}$ coincide with those in Ref.~\cite{Liao:2020roy}, but do not coincide with those in Ref.~\cite{Gonzalez:2015ady}. Hence the calculational details about the anomalous dimensions of $\mathcal O_4^{XY},\;\mathcal O_5^{XY}$ are given below:

\noindent$\bullet\mathcal{O}_4^{XX}:$
\begin{eqnarray}
&&{\rm (a)-(1)}:\frac{1}{4}C_F(\bar u\gamma_\alpha\gamma_\beta\gamma^\mu\gamma^\beta\gamma^\alpha P_Xd)(\bar u \sigma_{\mu\nu} P_Xd)=C_F(\bar u\gamma^\mu P_Xd)(\bar u\sigma_{\mu\nu} P_X d),\\
&&{\rm (a)-(2)}:\frac{1}{4}C_F(\bar u\gamma^\mu P_Xd)(\bar u\gamma_\alpha\gamma_\beta \sigma_{\mu\nu}\gamma^\beta\gamma^\alpha P_Xd)=0,\\
&&{\rm (b)-(1)}:-\frac{1}{4}(\bar u\gamma^\mu\gamma^\sigma\gamma^\alpha T^aP_X d)(\bar u \sigma_{\mu\nu}\gamma_\sigma\gamma_\alpha T^aP_X d)\nonumber\\
&&\qquad\quad\;\;=-3(\bar u\gamma^\mu T^aP_X d)(\bar u\sigma_{\mu\nu} T^aP_X d)+3i(\bar u\gamma_\nu T^aP_X d)(\bar u T^aP_X d),\label{Eq97}\\
&&(b)-(2):-\frac{1}{4}(\bar u\gamma^\sigma\gamma^\alpha\gamma^\mu T^aP_X d)(\bar u \gamma_\sigma\gamma_\alpha\sigma_{\mu\nu} T^aP_X d)\nonumber\\
&&\qquad\quad\;\;=-(\bar u\gamma^\mu T^aP_X d)(\bar u\sigma_{\mu\nu} T^aP_X d),\\
&&(c)-(1):\frac{1}{4}(\bar u\gamma^\mu\gamma^\sigma\gamma^\alpha T^aP_X d)(\bar u\gamma_\alpha\gamma_\sigma\sigma_{\mu\nu} T^aP_X d)\nonumber\\
&&\qquad\quad\;\;=3(\bar u\gamma^\mu T^aP_X d)(\bar u\sigma_{\mu\nu} T^aP_X d)+3i(\bar u\gamma_\nu T^aP_X d)(\bar u T^aP_X d),\\
&&(c)-(2):\frac{1}{4}(\bar u\gamma^\alpha\gamma^\sigma\gamma^\mu T^aP_X d)(\bar u\sigma_{\mu\nu}\gamma_\sigma\gamma_\alpha T^aP_X d)=(\bar u\gamma^\mu T^aP_X d)(\bar u\sigma_{\mu\nu} T^aP_X d),\\
&&\Rightarrow (a)+(b)+(c)=C_F(\bar u\gamma^\mu P_Xd)(\bar u\sigma_{\mu\nu} P_X d)-3(\bar u\gamma^\mu T^aP_X d)(\bar u\sigma_{\mu\nu} T^aP_X d)\nonumber\\
&&\qquad\qquad\qquad\qquad\;\;\;+3i(\bar u\gamma_\nu T^aP_X d)(\bar u T^aP_X d)-(\bar u\gamma^\mu T^aP_X d)(\bar u\sigma_{\mu\nu} T^aP_X d)\nonumber\\
&&\qquad\qquad\qquad\qquad\;\;\;+3(\bar u\gamma^\mu T^aP_X d)(\bar u\sigma_{\mu\nu} T^aP_X d)+3i(\bar u\gamma_\nu T^aP_X d)(\bar u T^aP_X d)\nonumber\\
&&\qquad\qquad\qquad\qquad\;\;\;+(\bar u\gamma^\mu T^aP_X d)(\bar u\sigma_{\mu\nu} T^aP_X d)\nonumber\\
&&\qquad\qquad\qquad\qquad=(C_F-\frac{3}{2})\mathcal{O}_4^{XX}+(-\frac{3}{2}i-\frac{3i}{N})\mathcal{O}_5^{XX}.\label{O4XX}
\end{eqnarray}

\noindent$\bullet\mathcal{O}_4^{XY}(X\neq Y):$
\begin{eqnarray}
&&{\rm (a)-(1)}:\frac{1}{4}C_F(\bar u\gamma_\alpha\gamma_\beta\gamma^\mu\gamma^\beta\gamma^\alpha P_Xd)(\bar u \sigma_{\mu\nu} P_Yd)=C_F(\bar u\gamma^\mu P_Xd)(\bar u\sigma_{\mu\nu} P_Y d),\\
&&{\rm (a)-(2)}:\frac{1}{4}C_F(\bar u\gamma^\mu P_Xd)(\bar u\gamma_\alpha\gamma_\beta \sigma_{\mu\nu}\gamma^\beta\gamma^\alpha P_Yd)=0,\\
&&{\rm (b)-(1)}:-\frac{1}{4}(\bar u\gamma^\mu\gamma^\sigma\gamma^\alpha T^aP_X d)(\bar u \sigma_{\mu\nu}\gamma_\sigma\gamma_\alpha T^aP_Y d)\nonumber\\
&&\qquad\quad\;\;=-(\bar u\gamma^\mu T^a P_Xd)(\bar u\sigma_{\mu\nu}T^a P_Y d),\\
&&{\rm (b)-(2)}:-\frac{1}{4}(\bar u\gamma^\sigma\gamma^\alpha\gamma^\mu T^aP_X d)(\bar u \gamma_\sigma\gamma_\alpha\sigma_{\mu\nu} T^aP_Y d)\nonumber\\
&&\qquad\quad\;\;=-3(\bar u\gamma^\mu T^a P_Xd)(\bar u\sigma_{\mu\nu}T^a P_Y d)-3i(\bar u\gamma_\nu T^a P_Xd)(\bar uT^a P_Y d),\\
&&{\rm (c)-(1)}:\frac{1}{4}(\bar u\gamma^\mu\gamma^\sigma\gamma^\alpha T^aP_X d)(\bar u\gamma_\alpha\gamma_\sigma\sigma_{\mu\nu} T^aP_Y d)=(\bar u\gamma^\mu T^a P_Xd)(\bar u\sigma_{\mu\nu}T^a P_Y d),\\
&&{\rm (c)-(2)}:\frac{1}{4}(\bar u\gamma^\alpha\gamma^\sigma\gamma^\mu T^aP_X d)(\bar u\sigma_{\mu\nu}\gamma_\sigma\gamma_\alpha T^aP_Y d)\nonumber\\
&&\qquad\quad\;\;=3(\bar u\gamma^\mu T^a P_Xd)(\bar u\sigma_{\mu\nu}T^a P_Y d)-3i(\bar u\gamma_\nu T^a P_Xd)(\bar uT^a P_Y d),\\
&&\Rightarrow (a)+(b)+(c)=C_F\mathcal{O}_4^{XY}-4i(\bar u\gamma_\nu T^a P_Xd)(\bar uT^a P_Y d)\nonumber\\
&&\qquad\qquad\qquad\quad\;\;\;=(C_F-\frac{3}{2})\mathcal{O}_4^{XY}+(\frac{3}{2}i+\frac{3i}{N})\mathcal{O}_5^{XY}.\label{O4XY}
\end{eqnarray}

\noindent$\bullet\mathcal{O}_5^{XX}:$
\begin{eqnarray}
&&{\rm (a)-(1)}:\frac{1}{4}C_F(\bar u\gamma_\alpha\gamma_\beta\gamma_\mu\gamma^\beta\gamma^\alpha P_Xd)(\bar u P_Xd)=C_F(\bar u\gamma^\mu P_Xd)(\bar u P_X d),\\
&&{\rm (a)-(2)}:\frac{1}{4}C_F(\bar u\gamma_\mu P_Xd)(\bar u\gamma_\alpha\gamma_\beta \gamma^\beta\gamma^\alpha P_Xd)=4C_F(\bar u\gamma^\mu P_Xd)(\bar u P_X d),\\
&&{\rm (b)-(1)}:-\frac{1}{4}(\bar u\gamma_\mu\gamma_\sigma\gamma_\alpha T^aP_X d)(\bar u \gamma^\sigma\gamma^\alpha T^aP_X d)\nonumber\\
&&\qquad\quad\;\;=-i(\bar u\gamma^\nu T^aP_X d)(\bar u \sigma_{\nu\mu} T^aP_X d)-(\bar u\gamma_\mu T^a P_Xd)(\bar u T^aP_X d),\\
&&{\rm (b)-(2)}:-\frac{1}{4}(\bar u\gamma_\sigma\gamma_\alpha\gamma_\mu T^aP_X d)(\bar u \gamma^\sigma\gamma^\alpha T^aP_X d)=-(\bar u\gamma_\mu T^a P_Xd)(\bar u T^aP_X d),\\
&&{\rm (c)-(1)}:\frac{1}{4}(\bar u\gamma_\mu\gamma_\sigma\gamma_\alpha T^aP_X d)(\bar u \gamma^\alpha\gamma^\sigma T^aP_X d)\nonumber\\
&&\qquad\quad\;\;=(\bar u\gamma_\mu T^a P_Xd)(\bar u T^aP_X d)-i(\bar u\gamma^\nu T^aP_X d)(\bar u \sigma_{\nu\mu} T^aP_X d),\label{Eq119}\\
&&{\rm (c)-(2)}:\frac{1}{4}(\bar u\gamma_\alpha\gamma_\sigma\gamma_\mu T^aP_X d)(\bar u \gamma^\sigma\gamma^\alpha T^aP_X d)=(\bar u\gamma_\mu T^a P_Xd)(\bar u T^aP_X d),\\
&&\Rightarrow (a)+(b)+(c)=5C_F\mathcal{O}_5^{XX}-2i(\bar u\gamma^\nu T^aP_X d)(\bar u \sigma_{\nu\mu} T^aP_X d)\nonumber\\
&&\qquad\qquad\qquad\quad\;\;\;=(5C_F-\frac{3}{2})\mathcal{O}_5^{XX}+(-\frac{i}{2}+\frac{i}{N})\mathcal{O}_4^{XX}.
\label{O5XX}
\end{eqnarray}

\noindent$\bullet\mathcal{O}_5^{XY}(X\neq Y):$
\begin{eqnarray}
&&{\rm (a)-(1)}:\frac{1}{4}C_F(\bar u\gamma_\alpha\gamma_\beta\gamma_\mu\gamma^\beta\gamma^\alpha P_Xd)(\bar u P_Yd)=C_F(\bar u\gamma^\mu P_Xd)(\bar u P_X d),\\
&&{\rm (a)-(2)}:\frac{1}{4}C_F(\bar u\gamma_\mu P_Xd)(\bar u\gamma_\alpha\gamma_\beta \gamma^\beta\gamma^\alpha P_Yd)=4C_F(\bar u\gamma^\mu P_Xd)(\bar u P_Y d),\\
&&{\rm (b)-(1)}:-\frac{1}{4}(\bar u\gamma_\mu\gamma_\sigma\gamma_\alpha T^aP_X d)(\bar u \gamma^\sigma\gamma^\alpha T^aP_Y d)=-(\bar u\gamma^\mu T^a P_Xd)(\bar u T^aP_Y d),\\
&&{\rm (b)-(2)}:-\frac{1}{4}(\bar u\gamma_\sigma\gamma_\alpha\gamma_\mu T^aP_X d)(\bar u \gamma^\sigma\gamma^\alpha T^aP_Y d)\nonumber\\
&&\qquad\quad\;\;=i(\bar u\gamma^\nu T^a\gamma^5P_X d)(\bar u \sigma_{\nu\mu} T^a\gamma^5P_Y d)-(\bar u\gamma^\mu T^a P_Xd)(\bar u T^aP_Y d),\label{Eq125}\\
&&{\rm (c)-(1)}:\frac{1}{4}(\bar u\gamma_\mu\gamma_\sigma\gamma_\alpha T^aP_X d)(\bar u \gamma^\alpha\gamma^\sigma T^aP_Y d)=(\bar u\gamma^\mu T^a P_Xd)(\bar u T^aP_Y d),\\
&&{\rm (c)-(2)}:\frac{1}{4}(\bar u\gamma_\alpha\gamma_\sigma\gamma_\mu T^aP_X d)(\bar u \gamma^\sigma\gamma^\alpha T^aP_Y d)\nonumber\\
&&\qquad\quad\;\;=i(\bar u\gamma^\nu T^a\gamma^5P_X d)(\bar u \sigma_{\nu\mu} T^a\gamma^5P_Y d)+(\bar u\gamma^\mu T^a P_Xd)(\bar u T^aP_Y d),\\
&&\Rightarrow (a)+(b)+(c)=5C_F\mathcal{O}_5^{XX}+2i(\bar u\gamma^\nu T^aP_X d)(\bar u \sigma_{\nu\mu} T^aP_X d)\nonumber\\
&&\qquad\qquad\qquad\quad\;\;\;=(5C_F-\frac{3}{2})\mathcal{O}_5^{XY}+(\frac{i}{2}-\frac{i}{N})\mathcal{O}_4^{XY}.
\label{O5XY}
\end{eqnarray}
The Fierz transformation formalisms used in above calculation read
\begin{eqnarray}
&&(P_X)_{12}(P_X)_{34}=\frac{1}{2}(P_X)_{14}(P_X)_{32}+\frac{1}{8}(\sigma_{\mu\nu}P_X)_{14}(\sigma^{\mu\nu}P_X)_{32},\nonumber\\
&&(P_X)_{12}(P_Y)_{34}=\frac{1}{2}(\gamma^\mu P_X)_{14}(\gamma_\mu P_Y)_{32},\nonumber\\
&&(\sigma_{\mu\nu}P_X)_{12}(\sigma^{\mu\nu}P_X)_{34}=6(P_X)_{14}(P_X)_{32}-
\frac{1}{2}(\sigma_{\mu\nu}P_X)_{14}(\sigma^{\mu\nu}P_X)_{32},\nonumber\\
&&(\gamma^\mu P_X)_{12}(\gamma_\mu P_X)_{34}=-(\gamma^\mu P_X)_{14}(\gamma_\mu P_X)_{32},\nonumber\\
&&(\gamma^\mu P_X)_{12}(\gamma_\mu P_Y)_{34}=2(P_X)_{14}(P_Y)_{32},\nonumber\\
&&(\gamma^\mu P_X)_{12}(P_X)_{34}=\frac{1}{2}(\gamma^\mu P_X)_{14}(P_X)_{32}-\frac{i}{2}(\gamma_\nu P_X)_{14}(\sigma^{\nu\mu} P_X)_{32},\nonumber\\
&&(\gamma^\mu P_X)_{12}(P_Y)_{34}=\frac{1}{2}(\gamma^\mu P_X)_{14}(P_Y)_{32}+\frac{i}{2}(\gamma_\nu P_X)_{14}(\sigma^{\nu\mu}P_Y)_{32},\nonumber\\
&&(\gamma_\mu P_X)_{12}(\sigma^{\mu\nu}P_X)_{34}=\frac{3i}{2}(\gamma^\nu P_X)_{14}( P_X)_{32}-\frac{1}{2}(\gamma_\mu P_X)_{14}(\sigma^{\mu\nu}P_X)_{32},\nonumber\\
&&(\gamma_\mu P_X)_{12}(\sigma^{\mu\nu}P_Y)_{34}=-\frac{3i}{2}(\gamma^\nu P_X)_{14}( P_X)_{32}-\frac{1}{2}(\gamma_\mu P_X)_{14}(\sigma^{\mu\nu} P_X)_{32},
\end{eqnarray}
and for the generators $T^a$ of $SU(N)$ we have
\begin{eqnarray}
&&T^a_{ij}T^a_{jk}=\frac{N^2-1}{2N}\delta_{ik}=C_F\delta_{ik},\;\;\;T^a_{ij}T^a_{kl}=\frac{1}{2}(\delta_{il}\delta_{kj}-
\frac{1}{N}\delta_{ij}\delta_{kl}).
\end{eqnarray}

Combining Eqs.~(\ref{A19}-\ref{QCD18}) with Eq.~(\ref{eq50}), we can compute out the numerical QCD RGE evolution matrices from $\Lambda=M_{W_L}$ to $\mu\simeq 1.0\;$GeV as
\begin{eqnarray}
&&\hat U_{(12)}^{XX}(\mu,\Lambda)=\left(\begin{array}{c}1.96 \;\;\quad 0.01\\ -2.82 \quad 0.45\end{array}\right),\;\;
\hat U_{(31)}^{XY}(\mu,\Lambda)=\left(\begin{array}{c}0.87 \quad -1.4\\ 0 \qquad 2.97\end{array}\right), \nonumber\\
&&U_3^{XX}(\mu,\Lambda)=0.76,\nonumber\\
&&\hat U_{(45)}^{XX}(\mu,\Lambda)=\left(\begin{array}{c}0.68 \qquad -0.24i\\ -0.016i \quad 0.71\end{array}\right),\;\;
\hat U_{(45)}^{XY}(\mu,\Lambda)=\left(\begin{array}{c}0.68 \qquad 0.34i\\ 0.023i \quad 1.4\end{array}\right).\label{48}
\end{eqnarray}

\end{appendix}

\end{document}